\documentclass[showpacs,preprintnumbers,amsmath,amssymb]{revtex4}
\usepackage{graphicx}% Include figure files
\usepackage{dcolumn}% Align table columns on decimal point
\usepackage{bm}% bold math
\usepackage{epsf}
\usepackage{subfigure}
\usepackage{epstopdf}
\usepackage{amsmath}
\usepackage{amssymb}
\newcommand{\bra}[1]{\langle #1|}
\newcommand{\ket}[1]{|#1\rangle}

\newcommand{\mean}[1]{\langle #1 \rangle}
\newcommand{\trace}{{\rm Tr}}

\newcommand{\sbra}[1]{\ll \hspace{-0.1cm} #1|}
\newcommand{\sket}[1]{| #1 \hspace{-0.1cm} \gg}
\newcommand{\sbraket}[2]{\ll \hspace{-0.1cm} #1|#2 \hspace{-0.1cm} \gg}

\begin{document}

\title{Fluctuation theorem for counting-statistics 
in electron transport through quantum junctions}

\author{Massimiliano Esposito}
\altaffiliation[Also at \; ]{Center for Nonlinear Phenomena and Complex Systems,
Universite Libre de Bruxelles, Code Postal 231, Campus Plaine,
B-1050 Brussels, Belgium.}
\author{Upendra Harbola}
\author{Shaul Mukamel}
\affiliation{Department of Chemistry, University of California,
Irvine, California 92697, USA.}

\date{\today}

\begin{abstract}
We demonstrate that the probability distribution of the net 
number of electrons passing through a quantum system in a 
junction obeys a steady-state fluctuation theorem (FT) which
can be tested experimentally by the full counting statistics 
(FCS) of electrons crossing the lead-system interface.
The FCS is calculated using a many-body quantum master equation (QME) 
combined with a Liouville space generating function (GF) formalism.
For a model of two coupled quantum dots, we show that the FT 
becomes valid for long binning times and provide an estimate 
for the finite-time deviations. 
We also demonstrate that the Mandel (or Fano) parameter associated 
with the incoming or outgoing electron transfers show 
subpoissonian (antibunching) statistics.
\end{abstract}

\maketitle
%%%%%%%%%%%%%%%%%%%%%%%%%%%%%%%%%%%%%%%%%%%%%%%%%%%%%%%%%%%%%%%%%%%%
\section{Introduction}

Various far-from-equilibrium relations, such as the Jarzynski relation
\cite{Jarzynski1,Crooks99,Crooks00} or the fluctuation theorem (FT)
\cite{Gallavotti,Kurchan1,Lebowitz,Searles,Gaspard1,Seifert,Broeck}, 
obtained for classical systems during the past decade, provide 
new insights into the emergence of irreversible processes 
in physical systems \cite{Maesrev,Gaspardrev}.
These relations follow from the observation that the ratio of the 
probability of a system forward and time-reversed trajectory
is given by the exponential of a quantity, the trajectory entropy 
production, which when ensemble averaged, gives the entropy production 
in the system.
These relations therefore quantify the probabilities of observing
"non-thermodynamic" trajectories with decreased trajectory entropy 
production.
These probabilities are infinitesimally small in the macroscopic world
but are non-negligible in microscopic systems.
However, the ensemble averaged dynamics always satisfies the second 
law (entropy production always grows).
With the recent progress in nano and mesoscopic sciences, 
these probabilities can now be measured \cite{Bustamente1,Bustamente2}.
Because in the microscopic world quantum effects can be important, it is
interesting to establish whether these fluctuation relations remain true in 
the quantum regime. This is still an open issue
\cite{MukamelQJ,STasaki,Maes,Monnai,Allahverdyan,Maes2,EspositoMukamel}.
One of the major obstacles for a general formulation of a quantum FT is
the lack of a clear concept of a measurable trajectory.
It is therefore helpful to consider systems undergoing a well-defined
measurement process.
In the counting statistics of photons emitted by an atom or a molecule
driven out of equilibrium by a laser field \cite{Glauber,Kelley,Mandel,
Brown,Mukamel03,Mukamel05,Mukamel06,BarkaiRev,Orrit}, a
trajectory picture is provided by the history of the detected photons.
However, the reverse trajectory (where the laser mode absorbs a photon
from the molecule) is not easily measurable.
Electron counting statistics provides on the other hand a clear trajectory 
picture given by the history of the electron transfers between the system 
and the leads, where the reverse trajectory 
(electron moving against the bias) is a measurable quantity.
Electron counting statistics in nanosystems has attracted recent interest
\cite{Levitov,Buttiker,Gurvitz,Rammer1,Rammer2,Rammer3,Jauho,Kießlich,
Utsumi,Pederson,Nazarov03}.
Individual electrons crossing quantum dots have been measured 
\cite{Lu,Fujisawa,Bylander,Gustavsson}.
Measuring the statistics of both forward and backward electron transfer 
events is essential to verify the FT and has recently been reported 
in Ref. \cite{Hirayama}.  
Most studies have focused on the few lowest moments of the distribution.
However, the FT is connected with the probabilities of large fluctuations 
which require the knowledge of the entire probability distribution.\\

In this paper, we use the many-electron QME derived in Ref.
\cite{HarbolaEsposito} and the generating operator (GO) formalism 
in Liouville space developed for photon counting statistics
\cite{Glauber,Kelley,Mandel,Brown,Mukamel03,Mukamel05,Mukamel06}
to calculate the FCS of electrons in biased quantum junctions. 
Our central formal result is an equation of motion for the GO whose solution 
can provide the full electron transfer probability distribution. 
Neglecting electron-electron interactions, this GO can be factorized 
into products of single orbital GO, each leading to a statistics 
similar to the one of the single resonant-level system studied in Ref.
\cite{Nazarov03}. 
By constructing the current GO from the full electron counting GO, we 
show that the probability distribution of the {\it net} number of 
electrons $k$ entering the system from one of the system-lead interface 
[$k(t)=- \frac{1}{e} \int_{0}^{t} d\tau I(\tau)$, where $I(\tau)$ is 
the current and $t$ the measurement time also called binning time], 
satisfies: 
$P_t(k) / P_t(-k) \stackrel{t \to \infty}{=} \exp{(\beta e V k)}$,
where $eV$ is the difference between the left and right 
lead chemical potentials and $\beta=(k_b T)^{-1}$ the inverse 
temperature.
The probability of observing a current in the direction favored by 
the bias voltage $V$ is exponentially larger than that of measuring 
the reverse current.
For a non-biased junction the two currents are equiprobable. \\

The QME presented in section \ref{QME} is used to calculate the 
FCS of electrons in section \ref{eleccont}.
In section \ref{CurrentGF}, we define the GO for the net number of 
electron transfer and show that the FT holds for long measurement times. 
In section \ref{AverageGF} we derive closed expressions for 
the current and its power spectrum and for the Mandel parameter.
In section \ref{twolevelmodel}, we calculate the probability 
distribution for the net number of electron transfer for a model of 
two-coupled quantum dots and analyze the finite-time deviations to the FT. 
We also study the behavior of the average 
current and Mandel parameter as a function of the bias and temperature.
Conclusions are drawn is section \ref{conclusion}. 

%%%%%%%%%%%%%%%%%%%%%%%%%%%%%%%%%%%%%%%%%%%%%%%%%%%%%%%%%%%%%%%%%%%%%%%%%%%%%%%%%%%%%%%%%%%%%%%%%%%%%%%%%%
\section{The Quantum master equation} \label{QME}

The quantum junction is made of a system (e.g. quantum dot or single 
molecule) coupled to two leads. 
The system Hamiltonian reads $H_S = \sum_{s} \epsilon_s c_s^\dag c_s$, 
where $c_s^\dag$ ($c_s$) is the Fermi creation (annihilation) operator 
for the $s$ system orbital. 
The Hamiltonian of the left (right) lead is  
$H_{L} = \sum_{l} \epsilon_{l} c_{l}^\dag c_{l}$  
($H_{R} = \sum_{r} \epsilon_{r} c_{r}^\dag c_{r}$), where
$l$ ($r$) runs over all the left (right) lead orbitals. 
The entire junction Hamiltonian is $H=H_S+H_L+H_R+H_T$ 
where $H_T =\sum_{s\nu} \left[T_{s\nu} c^\dag_s c_\nu + h.c. \right]$ 
is the system-leads coupling ($\nu=l,r$). 
In \cite{HarbolaEsposito}, we used second order perturbation theory in
the lead-system interaction and projection operators in the number
of electrons in the system to derive a QME describing the dynamics of
the reduced system density matrix. 
The QME gives a Redfield equation \cite{Breuer} in Fock space which only 
contains coherences between states with the same number of electrons. 
When the relaxation induced by the leads is much slower than the Bohr 
frequencies of the system, the fast oscillations in the interaction 
picture can be averaged out. 
This approximation, known as the rotating wave approximation (RWA) in 
quantum optics, is often performed on the Redfield equation to guarantee 
that the final QME is of the Lindblad form 
\cite{Breuer,Gardiner,Haake,Spohn,Tannoudji}. 
Our QME finally reads
\begin{eqnarray}
\dot{\rho}^n &=& -i[H_S,\rho^n]
+\sum_{s} \left(
v_{ss} c_{s} \rho^{n+1} c_s^\dag
+ w_{ss} c_s^\dag \rho^{n-1} c_{s}
- v_{ss} c_s^\dag c_{s} \rho^n - w_{ss} \rho^n c_{s} c_s^\dag + {\rm h.c.} \right)
\label{QMEaaaaaa} \;,
\end{eqnarray}
where $\rho^{n}$ is the reduced density matrix of the system projected
into the $n$ electron part of the Fock space.
The complete system density matrix is given by $\rho=\sum_n \rho^{n}$.
When summed over $n$, Eq. (\ref{QMEaaaaaa}) gives an equation for the 
total $\rho$ which is of the Lindblad form \cite{HarbolaEsposito}.
$v_{ss}$'s and $w_{ss}$'s are related to lead correlation functions.
Assuming a quasi-continuous spectra for the leads and neglecting the level shift
contributions (which only modify the bare Bohr frequencies of the system), we have
\begin{eqnarray}
v_{ss} \equiv \sum_y v_{ss}^{(y)} \  \ ; \  \
w_{ss} \equiv \sum_y w_{ss}^{(y)} \label{QMEaaaaab}
\end{eqnarray}
where
\begin{eqnarray}
v_{ss}^{(y)}
&=&\pi n_y(\epsilon_{s}) \vert T_{s}^{(y)}(\epsilon_{s}) \vert^2
(1-f_y(\epsilon_{s})) \label{QMEaaaaabb}\\
w_{ss}^{(y)}
&=&\pi n_y(\epsilon_{s}) \vert T_{s}^{(y)}(\epsilon_{s}) \vert^2
f_y(\epsilon_{s}) \nonumber \;.
\end{eqnarray}
$n_y(\epsilon)$ is the density of state of the left or right 
lead ($y=L,R$) at energy $\epsilon$.
$f_y(\epsilon) \equiv 1/(\exp{[\beta (\epsilon-\mu_y)}]+1)$ denotes
the Fermi distribution of the $y$ lead and $\mu_y$ is its chemical potentials.
We assume $\mu_R=\mu_0$ and $\mu_L=\mu_0+e V$, where $V$ is the 
bias [see Fig.\ref{fig1}].
$w_{ss}^{(y)}$ is the electron transfer rate from lead $y$ to 
the $s$ orbital and $v_{ss}^{(y)}$ is the rate for the reverse process.
These obey the relation
\begin{eqnarray}
v_{ss}^{(y)} =
e^{\beta (\epsilon_{s}-\mu_{y})} w_{ss}^{(y)} \;,
\label{QMEaaaaac}
\end{eqnarray}
so that
\begin{eqnarray}
\frac{v_{ss}^{(R)} w_{ss}^{(L)}}
{v_{ss}^{(L)} w_{ss}^{(R)}} = e^{\beta e V} \;.
\label{QMEaaaaad}
\end{eqnarray}
%This relation will be crucial in proving the FT.

%%%%%%%%%%%%%%%%%%%%%%%%%%%%%%%%%%%%%%%%%%%%%%%%%%%%%%%%%%%%%%%%%%%%%%%%%%%%%%%%%%%%%%%%%%%%%%%%%%%%%%%%%%
\section{Generating-function for electron-counting statistics} \label{eleccont}

We consider a system with $M$ orbitals and
$n$ spinless electrons, so that $n=0,1,\ldots,M$.
The number of $n$-electron many-body states (hereafter denoted states) 
is given by $C^{M}_{n}=\frac{M!}{(M-n)!n!}$. 
The total number of Fock space states is 
$N_{\rm tot}=\sum_{n=0}^{M} C^{M}_{n}=2^M$. 
As a result of the weak lead-system coupling and infinite leads 
assumption, the Fock space coherences (FSC) between many-body states 
with different $n$ are neglected and the number of elements of the full 
many-body density matrix reduces from $N_{\rm tot}^2=4^{M}$ to
$N_{\rm red}^2=\sum_{n=0}^{M} (C^{M}_{n})^2$. 
The space of the density matrices where FSC have 
been eliminated constitute our reduced Liouville space. 
By expanding the QME (\ref{QMEaaaaaa}) in the eigenbasis of the 
system, the population dynamics obeys a birth and death master 
equation which is decoupled from the coherence dynamics.
Electron-transfer events are counted by identifying the
terms in the QME which are responsible for the transitions between 
the populations. Their sequence constitutes a "trajectory".\\

We shall recast the QME (\ref{QMEaaaaaa}) in our reduced Liouville space as
\begin{eqnarray}
\sket{\dot{\rho}} 
= \big( \hat{{\cal L}} + \hat{\gamma} + \hat{\Gamma} \big) \sket{\rho}
\equiv \hat{{\cal M}} \sket{\rho} \;.
\label{ecountAaaaa}
\end{eqnarray}
$\hat{{\cal M}}$ is the generator of the QME. 
$\hat{{\cal L}}$ describes the isolated system dynamics
\begin{eqnarray}
\hat{{\cal L}} &=&
-i \sum_{s} \epsilon_s
[c_s^\dag c_{s}, \cdot ] \;. \label{ecountAaaaf}
\end{eqnarray}
We denote the four possible processes depicted in 
Fig. \ref{fig1} by $\eta=1,2,3,4$. 
$\eta=1$ and $\eta=3$ represent electron transfer from the system 
to the left and right lead whereas $\eta=2$ and $\eta=4$ represent 
the electron transfer from the left and right lead to the system. 
The orbital through which electron transfer occurs is denoted by $s$.
$\hat{\Gamma}$ is responsible for electron transfers and is made of 
the non-diagonal terms of the generator which couple the populations
\begin{eqnarray}
\hat{\Gamma} &=& \sum_{\nu} \hat{\Gamma}_{\nu} \;,
\label{ecountAaaab}
\end{eqnarray}
where $\nu=(\eta,s)$, $\sum_{\nu}=\sum_{\eta=1}^{4} \sum_{s=1}^{M}$,
and
\begin{eqnarray}
\hat{\Gamma}_{(1,s)} &\equiv&
2 v_{ss}^{(L)} c_s \cdot c_{s}^\dag \ \ ; \ \
\hat{\Gamma}_{(2,s)} \equiv
2 w_{ss}^{(L)} c_s^\dag \cdot c_{s} \nonumber \\
\hat{\Gamma}_{(3,s)} &\equiv&
2 v_{ss}^{(R)} c_s \cdot c_{s}^\dag \ \ ; \ \
\hat{\Gamma}_{(4,s)} \equiv
2 w_{ss}^{(R)} c_s^\dag \cdot c_{s} \:.
\label{ecountAaaac}
\end{eqnarray}
$\hat{\gamma}$ describes the diagonal terms of the generator 
\begin{eqnarray}
\hat{\gamma} &=& \sum_{\nu} \hat{\gamma}_{\nu} \;.
\label{ecountAaaad}
\end{eqnarray}
In analogy to $\hat{\Gamma}$, we identify the various contributions 
to $\hat{\gamma}$ from the different active orbitals
\begin{eqnarray}
\hat{\gamma}_{(1,s)} &\equiv&
- v_{ss}^{(L)} [c_s^\dag c_{s}, \cdot]_+  \ \ ; \ \
\hat{\gamma}_{(2,s)} \equiv
w_{ss}^{(L)} [c_s c_{s}^\dag, \cdot]_+ \nonumber \\
\hat{\gamma}_{(3,s)} &\equiv&
- v_{ss}^{(R)} [c_s^\dag c_{s}, \cdot]_+ \ \ ; \ \
\hat{\gamma}_{(4,s)} \equiv
w_{ss}^{(R)} [c_s c_{s}^\dag, \cdot]_+ \; .
\label{ecountAaaae}
\end{eqnarray}

We can now calculate the full electron counting statistics using
the formalism developed for photon counting statistics
\cite{Glauber,Kelley,Mandel,
Brown,Mukamel03,Mukamel05,Mukamel06,BarkaiRev,Orrit}.
Starting with a trajectory picture of the QME evolution
in terms of electron transfer histories, we shall calculate the 
probabilities of these trajectories and their associated generating 
function (GF). \\
The system density matrix conditional to measuring $\bold{k}$ electron 
transfers during an interval of time $t$ is denoted $\rho^{(\bold{k})}(t)$. 
We use the compact notation defined in appendix \ref{trajectories} 
[see (\ref{ecountAaaaob1})]. 
$\bold{k}$ is a vector with components $k_{\nu}$. 
The probability to measure $\bold{k}$ electron transfers during a time 
interval $t$ is obtained by tracing the conditional system density matrix
\begin{eqnarray}
P(t,\bold{k})
= \sbraket{I}{\rho^{(\bold{k})}(t)} \;.
\label{ecountAaaaqamain}
\end{eqnarray}
The trace of an Hilbert space operator $A$, $\trace A$, is denoted 
as a scalar product in Liouville space $\sbraket{I}{A}$, where $I$ 
is the unity operator.\\

The generating function (GF) associated with this probability 
distribution is defined as
\begin{eqnarray}
G(t,\boldsymbol{\lambda}) &\equiv&
\sum_{\bold{k}} P(t,\bold{k}) e^{\boldsymbol{\lambda} \cdot \bold{k}} \;,
\label{ecountAaaara}
\end{eqnarray}
where $\boldsymbol{\lambda} \cdot \bold{k}=\sum_{\nu} \lambda_{\nu} k_{\nu}$.
Similarly, we define the generating operator (GO) as
\begin{eqnarray}
\sket{G(t,\boldsymbol{\lambda})} &\equiv&
\sum_{\bold{k}} \sket{\rho^{(\bold{k})}(t)} e^{\boldsymbol{\lambda} \cdot \bold{k}} 
\;. \label{ecountAaaarc}
\end{eqnarray}
The GF is obtained by tracing the GO
\begin{eqnarray}
G(t,\boldsymbol{\lambda}) = \sbraket{I}{G(t,\boldsymbol{\lambda})} \;.
\label{ecountAaaarcamain}
\end{eqnarray}
An evolution equation for the GO is derived in appendix 
\ref{trajectories} starting with the QME
\begin{eqnarray}
\sket{\dot{G}(t,\boldsymbol{\lambda})}= \hat{{\cal W}}(\boldsymbol{\lambda}) 
\sket{G(t,\boldsymbol{\lambda})} \;,
\label{ecountAaaatab}
\end{eqnarray}
where 
\begin{eqnarray}
\hat{{\cal W}}(\boldsymbol{\lambda}) = \hat{{\cal M}}
+ \sum_{\nu} (e^{\lambda_{\nu}}-1) \hat{\Gamma}_{\nu}
\label{ecountAaaatac}
\end{eqnarray}
is the generator of the GO evolution equation.
Using the initial condition $\sket{G(0;\boldsymbol{\lambda})}=\sket{\rho(0)}$,
the solution of (\ref{ecountAaaatab}), given in appendix \ref{singlebody},
provides the GF at all times
\begin{eqnarray}
G(t,\boldsymbol{\lambda}) = \sbra{I} e^{ \hat{{\cal W}}(\boldsymbol{\lambda}) t}
\sket{\rho(0)} \;.\label{ecountAaaatc}
\end{eqnarray}
The GF contains the entire information about the electron
counting statistics. 
The probability distribution is obtained by inverting 
Eq. (\ref{ecountAaaara})
\begin{eqnarray}
P(t,\bold{k}) = \int_{0}^{2 \pi} d\boldsymbol{\lambda} \;
G(t,i \boldsymbol{\lambda}) e^{-i \boldsymbol{\lambda} \cdot \bold{k}} \;.
\label{ecountAaaarcb}
\end{eqnarray}
Moments of the distribution are given by derivatives 
of the GF
\begin{eqnarray}
\left. \frac{\partial^{n_1}}{\partial\lambda_{\nu_1}^{n_1}}
\frac{\partial^{n_2}}{\partial \lambda_{\nu_2}^{n_2}} \cdots
\frac{\partial^{n_N}}{\partial \lambda_{\nu_N}^{n_N}} 
G(t,\boldsymbol{\lambda})
\right|_{\boldsymbol{\lambda}=0} = \mean{k_{\nu_1}^{n_1}
k_{\nu_2}^{n_2} \cdots k_{\nu_N}^{n_N}}_t \;.
\label{ecountAaaarcc}
\end{eqnarray}
We also define 
\begin{eqnarray}
S(\boldsymbol{\lambda}) \equiv -\lim_{t \to \infty} 
\frac{1}{t} \ln G(t,\boldsymbol{\lambda}) 
\;. \label{FTaaaaae}
\end{eqnarray}
This will be useful to calculate the statistical 
properties of steady-state currents.

%%%%%%%%%%%%%%%%%%%%%%%%%%%%%%%%%%%%%%%%%%%%%%%%%%%%%%%%%%%%%%%%%%%%%%%%%%%%%%%%%%%%%%%%%%%%%%%%%%%%%%
\section{Fluctuation theorem for the net number of electrons transferred} 
\label{CurrentGF}

We will now focus on the statistical properties of the charge 
currents across the junction. 
We adopt the standard convention that the direction of the charge 
current is opposite to the electron transfers [see Fig.(\ref{fig1})]. 
The number of electrons transferred via process $\eta$ through 
orbital $s$ during a time interval $t$ is given by
\begin{eqnarray}
k_{\nu}(t) = - \frac{1}{e} 
\int_{0}^{t} d\tau I_{\nu}(\tau) \;,
\label{Caaaab}
\end{eqnarray}
where $I_{\nu}(\tau)$ is the corresponding charge current [$\nu=(\eta,s)$].
The net number of electron transfer events between the left (right) 
lead-system interface, through orbital $s$, during time $t$ is 
\begin{eqnarray}
k_{(L,s)}(t) \equiv k_{(2,s)}(t) - k_{(1,s)}(t) 
=- \frac{1}{e} \int_{0}^{t} d\tau I_{(L,s)}(\tau) \  \  \;, \  \
k_{(R,s)}(t) \equiv k_{(3,s)}(t) - k_{(4,s)}(t) 
=- \frac{1}{e} \int_{0}^{t} d\tau I_{(R,s)}(\tau) \;. \label{Caaaad}
\end{eqnarray}
where the charge current at the left [right] lead-system interface 
passing through the $s$'th system orbital is given by
$I_{(L,s)}(t) \equiv I_{(2,s)}(t) - I_{(1,s)}(t)$ 
[$I_{(R,s)}(t) \equiv I_{(3,s)}(t) - I_{(4,s)}(t)$].
The GF associated with the left and right net number of electron transfer 
$G(t,\boldsymbol{\lambda}_L,\boldsymbol{\lambda}_R)$
is the GF of the FCS $G(t,\boldsymbol{\lambda})$ where
$\lambda_{1,s}=-\lambda_{2,s} \equiv \lambda_{L,s}$
and $\lambda_{4,s}=-\lambda_{3,s} \equiv \lambda_{R,s}$. 
Defining the vectors $\boldsymbol{\lambda}_L$ ($\boldsymbol{\lambda}_R$) 
with components $\lambda_{L,s}$ ($\lambda_{R,s}$), we have
\begin{eqnarray}
G(t,\boldsymbol{\lambda}_L,\boldsymbol{\lambda}_R)
&=& \sum_{\boldsymbol{\lambda}_L,\boldsymbol{\lambda}_R} 
P(t,\bold{k}_L,\bold{k}_R)
e^{- \boldsymbol{\lambda}_L \cdot \bold{k}_L }
e^{- \boldsymbol{\lambda}_R \cdot \bold{k}_R } \;.
\label{Caaaaea}
\end{eqnarray}
For clarity, we hereafter consider the left GF. 
The right one may be calculated similarly.
The GF for the left net number of electron 
transfer $G(t,\boldsymbol{\lambda}_L)$ is defined from 
(\ref{Caaaaea}) by taking $\boldsymbol{\lambda}_R=0$.
Using Eq. (\ref{singleAaaa6a}), we find that
the generator $\hat{{\cal W}}_s(\lambda_{(L,s)})$ 
of the evolution of the single orbital GF $G_s(t,\lambda_{(L,s)})$ 
for the net number of electron transfer, via the single orbital $s$, 
through the left lead-system interface is given by 
\begin{eqnarray}
G_s(t,\lambda_{(L,s)}) &=& 
c_+(0) e^{g_{+}(\lambda_{(L,s)}) t}
\big( g^{+}_{1;1}(\lambda_{(L,s)}) +
g^{+}_{0;0}(\lambda_{(L,s)}) \big) \nonumber \\
&&+ c_{-}(0) e^{g_{-}(\lambda_{(L,s)})t}
\big( g^{-}_{1;1}(\lambda_{(L,s)}) +
g^{-}_{0;0}(\lambda_{(L,s)}) \big) \;,
\label{Caaaag}
\end{eqnarray}
where, using (\ref{singleAaaa6}) and (\ref{QMEaaaaad}), 
the eigenvalues of the generator are given by
\begin{eqnarray}
g_{\pm}(\lambda_{(L,s)}) &=& -\left(\frac{v_{ss}+w_{ss}}{2}\right) \pm
\sqrt{\left(\frac{v_{ss}+w_{ss}}{2}\right)^2 
+ v_{ss}^{(R)} w_{ss}^{(L)} \left[  e^{-\beta e V} (e^{\lambda_{(L,s)}}-1) 
+ (e^{-\lambda_{(L,s)}}-1) \right]} \;.
\label{Caaaah}
\end{eqnarray}
These possess the symmetry 
$g(\lambda_{(L,s)}) = g(\beta e V-\lambda_{(L,s)})$ so that,
using $S_s(\lambda_{(L,s)}) \equiv -\lim_{t \to \infty} 
\frac{1}{t} \ln G_s(t,\lambda_{(L,s)})$ and (\ref{Caaaag}), 
\begin{eqnarray}
S_s(\lambda_{(L,s)}) = S_s(\beta e V-\lambda_{(L,s)}) \;.
\label{Caaaai}
\end{eqnarray}
In appendix \ref{singlebody}, we show that the many-body GF can be 
factorized into a product of single orbital GF [see (\ref{singleAaaa7})].
The GF for the total current (irrespective of the carrying orbitals) by 
setting $\lambda_{(L,s)}=\lambda_L$ in $\boldsymbol{\lambda}_L$) 
can therefore be written
\begin{eqnarray}
G(t,\lambda_L) = 
\sum_{m}^{M} e^{g_{m}(\lambda_L) t}
\sbraket{I}{g_{m}(\lambda_L)} \sbraket{\tilde{g}_m(\lambda_L)}{\rho(0)} 
\;, \label{Caaaaia}
\end{eqnarray}
where $g_{m}(\lambda_L)$, $\sket{g_{m}(\lambda_L)}$ and
$\sbra{\tilde{g}_{m}(\lambda_L)}$ are respectively the many-body
eigenvalues, the right and the left eigenvector of the generator. 
Since the many-body eigenvalues corresponding to populations are 
made of $2^M$ possible sums of single-body orbital eigenvalues 
(\ref{Caaaah}), they also satisfy the symmetry
$g(\lambda_L) = g(\beta e V-\lambda_L)$ so that, using (\ref{FTaaaaae}),
\begin{eqnarray}
S(\lambda_L) = S(\beta e V-\lambda_L) \;.
\label{Caaaaib}
\end{eqnarray}
An important point is that the eigenvalues of the generator associated 
with the right GF are the same as those of the left GF so 
that $S(\lambda_L)=S(\lambda_R)$. 
The eigenvectors will however be different and 
$G(t,\lambda_L) \neq G(t,\lambda_R)$.
This means that, in general, the electron transfer statistics
at the left and right interface of the junction can be different.
However, the statistical properties which can be obtained from 
$S(\lambda)$ are the same on the two interface.
We will see that these include for example the FT 
[which follows from (\ref{Caaaaib})], the steady state average 
current [see (\ref{momentAaaaaab})], the current power spectrum 
at zero frequency [see (\ref{momentAaaaae})] or the asymptotic 
value of the Mandel parameter [see (\ref{spectdecAaaaalbis})].
Hereafter, we therefore omit the $L,R$ labeling in the 
corresponding quantities.\\

In appendix \ref{largedev}, we use the theory of large deviations 
to show that the symmetry (\ref{Caaaaib}) implies at long times 
\begin{eqnarray}
\frac{P(t,k)}{P(t,-k)} \stackrel{t \to \infty}{=} e^{\beta e V k}
\;. \label{FTaaaaah}
\end{eqnarray}
This is the FT for the net number of electrons $k$ 
crossing the junction at each system-lead interface.
Using (\ref{Caaaai}), we note that the FT also holds for the net number 
of charges which passed through each orbital $k_{(y,s)}$ ($y=L,R$). 
A similar result was pointed out in Ref. \cite{GaspardAndrieux2} 
for a single resonant level in the large Coulomb repulsion limit 
excluding double occupancy.
The FT implies that measuring electron transfers in the direction favored 
by the bias is exponentially more probable than the reverse process.
The argument of the exponential is proportional to the nonequilibrium 
constrains of the junction $\beta e V$ so that at equilibrium ($V=0$) 
the two probabilities are identical.

%%%%%%%%%%%%%%%%%%%%%%%%%%%%%%%%%%%%%%%%%%%%%%%%%%%%%%%%%%%%%%%%%%%%%%%%%%%%%%%%%%%%%%%%%%%%%%%%%%%%%%
\section{Average current, Mandel parameter and power spectrum} 
\label{AverageGF}

In appendix \ref{momentsappendix}, we show how currents, 
moments and cumulants can be obtained from the GF.
Using these results and the expressions for the eigenvalue with the 
smallest absolute value of the GF generators (\ref{singleAaaa6}),
we derive closed expressions for the steady state 
currents, their zero frequency power spectrum and the 
asymptotic value of the Mandel parameter.\\
 
Using (\ref{momentAaaaaab}) and (\ref{singleAaaa6}), the four steady 
state average currents through orbital $s$ are given by
\begin{eqnarray}
\mean{I_{(1,s)}}_{\rm st} &=& - 2 e v_{ss}^{(L)}
\frac{w_{ss}}{v_{ss}+w_{ss}} \; \ \ \;, \  \ \; 
\mean{I_{(2,s)}}_{\rm st} = - 2 e w_{ss}^{(L)}
\frac{v_{ss}}{v_{ss}+w_{ss}} \nonumber\\
\mean{I_{(3,s)}}_{\rm st} &=& - 2 e v_{ss}^{(R)}
\frac{w_{ss}}{v_{ss}+w_{ss}} \; \ \ \;, \  \ \; 
\mean{I_{(4,s)}}_{\rm st} = - 2 e w_{ss}^{(R)}
\frac{v_{ss}}{v_{ss}+w_{ss}} \;.\label{Caaaaj}
\end{eqnarray}
The total current associated with the process $\eta$ 
is obtained by summing over the orbitals
$\mean{I_{\eta}}_{\rm st} = \sum_{s}^{M} \mean{I_{(\eta,s)}}_{\rm st}$.
The Fano parameter $F_{\nu}(t)$ and the closely related Mandel 
parameter $M_{\nu}(t)$ of each of these processes $\nu=(\eta,s)$
are defined as
\begin{eqnarray}
F_{\nu}(t) \equiv M_{\nu}(t) + 1  \equiv \frac{\mean{k_{\nu}^2} -
\mean{k_{\nu}}_t^2}{\mean{k_{\nu}}_t} \label{Caaaajb}\;.
\end{eqnarray}
The Mandel parameter vanishes for a Poisson process.
$M<0$ ($M>0$) implies subpoissonian (superpoissonian) statistics.
At steady state, using (\ref{spectdecAaaaal}), we find 
\begin{eqnarray}
M_{\nu}(\infty) = \frac{1}{e}
\frac{\mean{I_{\nu}}_{\rm st}}{v_{ss}+w_{ss}} \;.
\label{Caaaak}
\end{eqnarray}
Another quantity of interest is the zero frequency power 
spectrum of the current $A_{\nu \nu}$ associated with 
the processes $\nu$ [see Eq. (\ref{momentAaaaae})]. 
$M_{\nu}(t)$ and $F_{\nu}(t)$ are easily related to it by 
$A_{\nu \nu} = e \mean{I_{\nu}}_{\rm st} F_{\nu}(\infty)$.
The zero frequency power spectrum for the 
total current associated with the process $\eta$ is 
given by $A_{\eta} = \sum_{s}^{M} A_{\nu \nu}$.  
The Mandel or Fano parameters cannot be expressed 
as such an orbital sum.
It is therefore convenient to calculate them from 
$S_{\eta}$ and $\mean{I_{\eta}}_{\rm st}$.\\
Similarly as for the $\eta$ processes, we can use (\ref{Caaaah}) 
to calculate the statistical properties of a given junction interface.
Using (\ref{momentAaaaaab}), the average steady state current 
via the s orbital reads
\begin{eqnarray}
\mean{I_{s}}_{\rm st} = \mean{I_{(R,s)}}_{\rm st} 
= 2 e \bigg(\frac{ v_{ss}^{(L)} w_{ss}^{(R)} 
- v_{ss}^{(R)} w_{ss}^{(L)}}{v_{ss}+w_{ss}} \bigg)
\;.\label{Caaaakc}
\end{eqnarray}
Using (\ref{momentAaaaae}), the corresponding zero frequency power 
spectrum of the current is given by
\begin{eqnarray}
A_{ss} = \frac{\mean{I_{s}}_{\rm st}^2}{v_{ss}+w_{ss}}
- 2 e^2 \bigg(\frac{v_{ss}^{(L)} w_{ss}^{(R)} 
+ v_{ss}^{(R)} w_{ss}^{(L)}}{v_{ss}+w_{ss}}\bigg) \;,
\label{Caaaakb}
\end{eqnarray}
and $A = \sum_{s}^{M} A_{ss}$.

%%%%%%%%%%%%%%%%%%%%%%%%%%%%%%%%%%%%%%%%%%%%%%%%%%%%%%%%%%%%%%%%%%%%%%%%%%%%%%%%%%%%%%%%%%%%%%%%%%
\section{Two coupled quantum dot model} \label{twolevelmodel}

\begin{figure}[p]
\centering
\rotatebox{0}{\scalebox{0.8}{\includegraphics{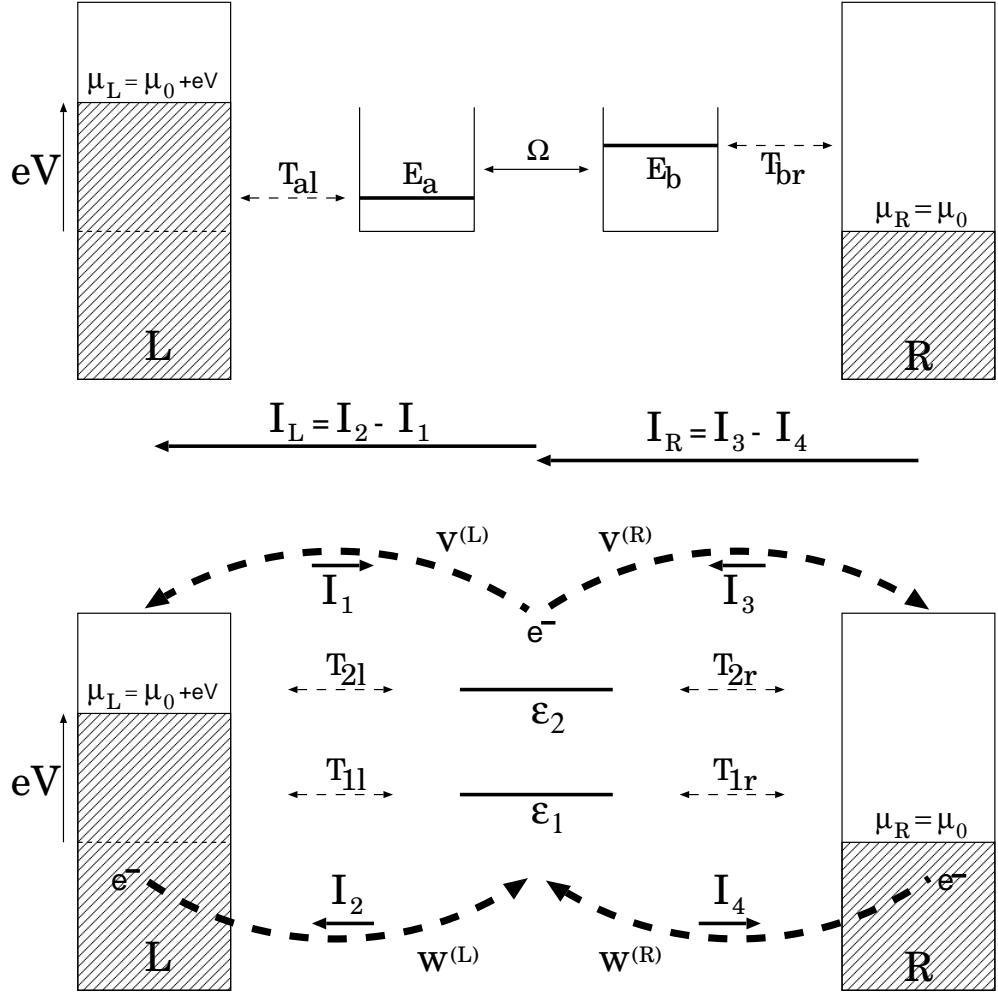}}}
\caption{Schematic representation of the model of two quantum dots $a$ 
and $b$ coupled in series between two leads used for our numerical results. 
The upper part depicts the system in the local basis where the Hamiltonian 
of the dots reads $H_S = \sum_{i,j} H_{ij} c_i^\dag c_j$,
where $H_{aa}=E_a=2$, $H_{bb}=E_b=5$ and $H_{ab}=H_{ba}^*=\Omega=1$.
The coupling element with the leads are $T_{al}=0.5$ $T_{br}=0.3$ and
$T_{ar}=T_{bl}=0$. 
The lower part depicts the system in the eigenbasis where the 
Hamiltonian becomes $H_S = \sum_{s} \epsilon_s c_s^\dag c_s$, 
with $\epsilon_1=1.697$ and $\epsilon_2=5.303$.
The coupling element with the leads become
$T_{1r}=-0.479$, $T_{2l}=0.145$, $T_{1r}=0.087$, $T_{2r}=0.287$.
We choose $\mu_0=0$ and $n_y(\epsilon_s)=\pi^{-1}$ as well as
$e=1$, $\hbar=1$ and $k_b=1$, so that the units of energy 
and temperature is $\Omega$ and the time unit $\Omega^{-1}$.
The four possible types of electron transfer (and their associated 
currents) are represented by the big dashed arrow.}
\label{fig1}
\end{figure}

We have calculated the probability distribution of the net 
number of electron transfer at the left lead-system interface 
for a model of two coupled quantum dots $a$ and $b$.
Quantities in local basis will be denoted by a tilde.
The Hamiltonian of the dots in the local basis reads
\begin{eqnarray}
\tilde{H}_s=\left(
\begin{array}{cc}
\epsilon_a & \Omega \\
\Omega & \epsilon_b
\end{array}
\right)
\end{eqnarray}
In the orbital eigenbasis, $H_s$ is a diagonal matrix
with eigenvalues
\begin{eqnarray}
\epsilon_{1,2}=\frac{\epsilon_a+\epsilon_b}{2}
\pm \sqrt{ (\frac{\epsilon_a-\epsilon_b}{2})^2 + \Omega^2 }
\end{eqnarray}
The orbitals are labeled by $s=1,2$.
The two Hamiltonians are connected by a unitary transformation 
$H_s = U \tilde{H}_s U^{\dagger}$, and similarly 
$\alpha=U \tilde{\alpha} U^{\dagger}$ and $\beta=U \tilde{\beta} U^{\dagger}$.
%\begin{eqnarray}
%U = \left(
%\begin{array}{cc}
%c_1  & \frac{c_2 (\epsilon_2-\epsilon_b)}{\Omega} \\
%\frac{c_1(\epsilon_1-\epsilon_a)}{\Omega} & c_2
%\end{array}
%\right)
%\end{eqnarray}
%where $c_1=\left(\sqrt{1+(\frac{\Omega}{\epsilon_1-\epsilon_a})^2}\right)^{-1}$ 
%and $c_2=\left(\sqrt{1+(\frac{\Omega}{\epsilon_2-\epsilon_b})^2}\right)^{-1}$.
We define the couplings between the leads and the dots in the local
basis and transform them to the orbital eigenbasis using $U$ [see Fig. \ref{fig1}].
Since the two orbitals can be either empty or singly occupied, the system 
has $4$ many-body states $\ket{00},\ket{01},\ket{10},\ket{11}$.
The many-body density matrix in the full Liouville space is thus 
a vector with $16$ elements ($4$ populations and $12$ coherences).
In our reduced Liouville space it is a vector with $6$ elements
($4$ populations and $2$ coherences between $\ket{01}$ and $\ket{10}$).
The generator (\ref{ecountAaaatac}) for this 
model is given in appendix \ref{twolevel}.\\

We have solved the GO equation (\ref{ecountAaaatab}) for the total left 
current, trace the solution (\ref{Caaaaia}) to get the GF, and finally 
calculate the probability distribution using (\ref{ecountAaaarcb}). 
We assume that the measurement starts when the junction is at steady state.
The parameters used in the numerical simulation 
are given in the legend of Fig. \ref{fig1}.\\

\begin{figure}[p]
\centering
\begin{tabular}{c}
\vspace{0.5cm}
\rotatebox{0}{\scalebox{0.42}{\includegraphics{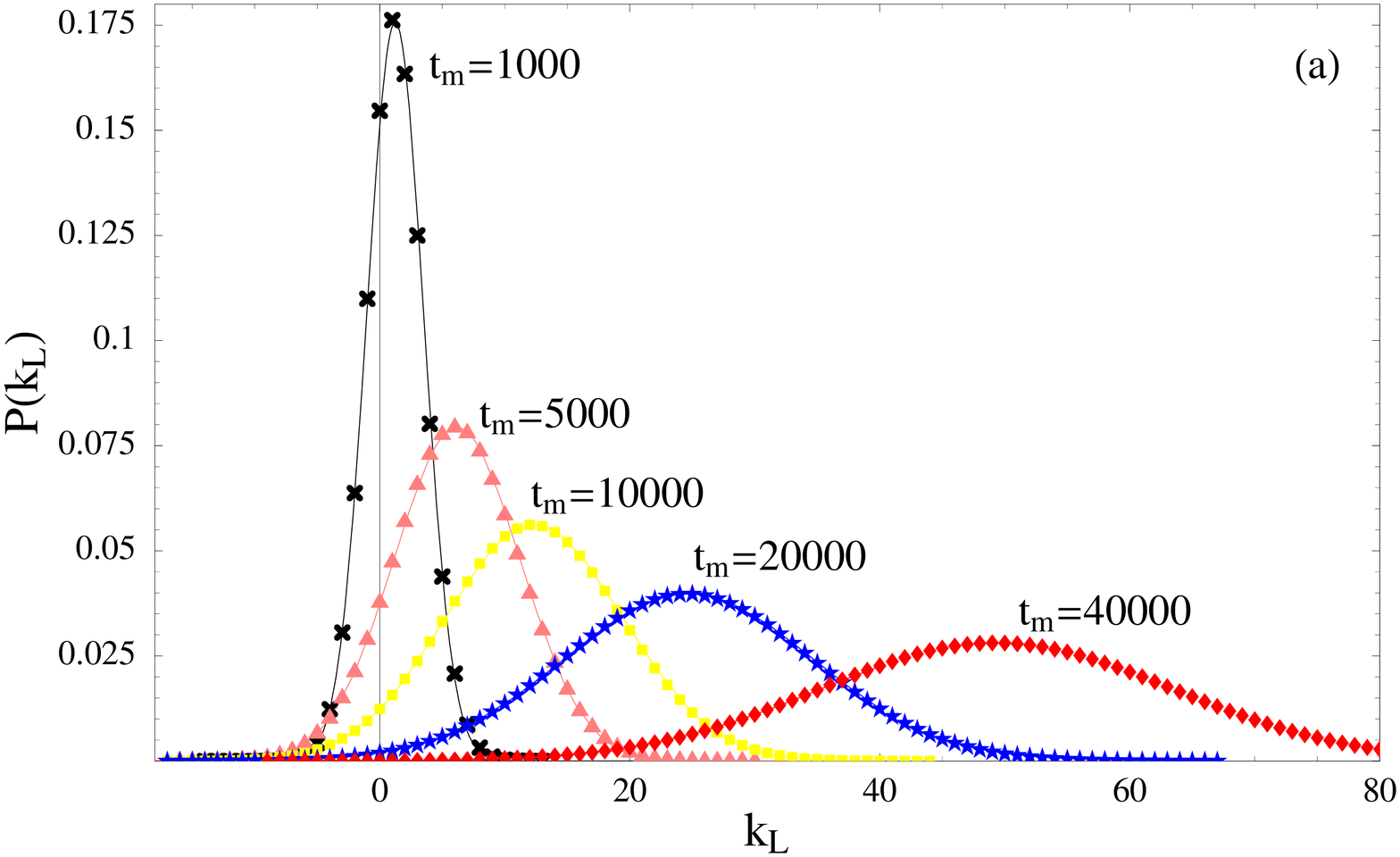}}}\\
\rotatebox{0}{\scalebox{0.42}{\includegraphics{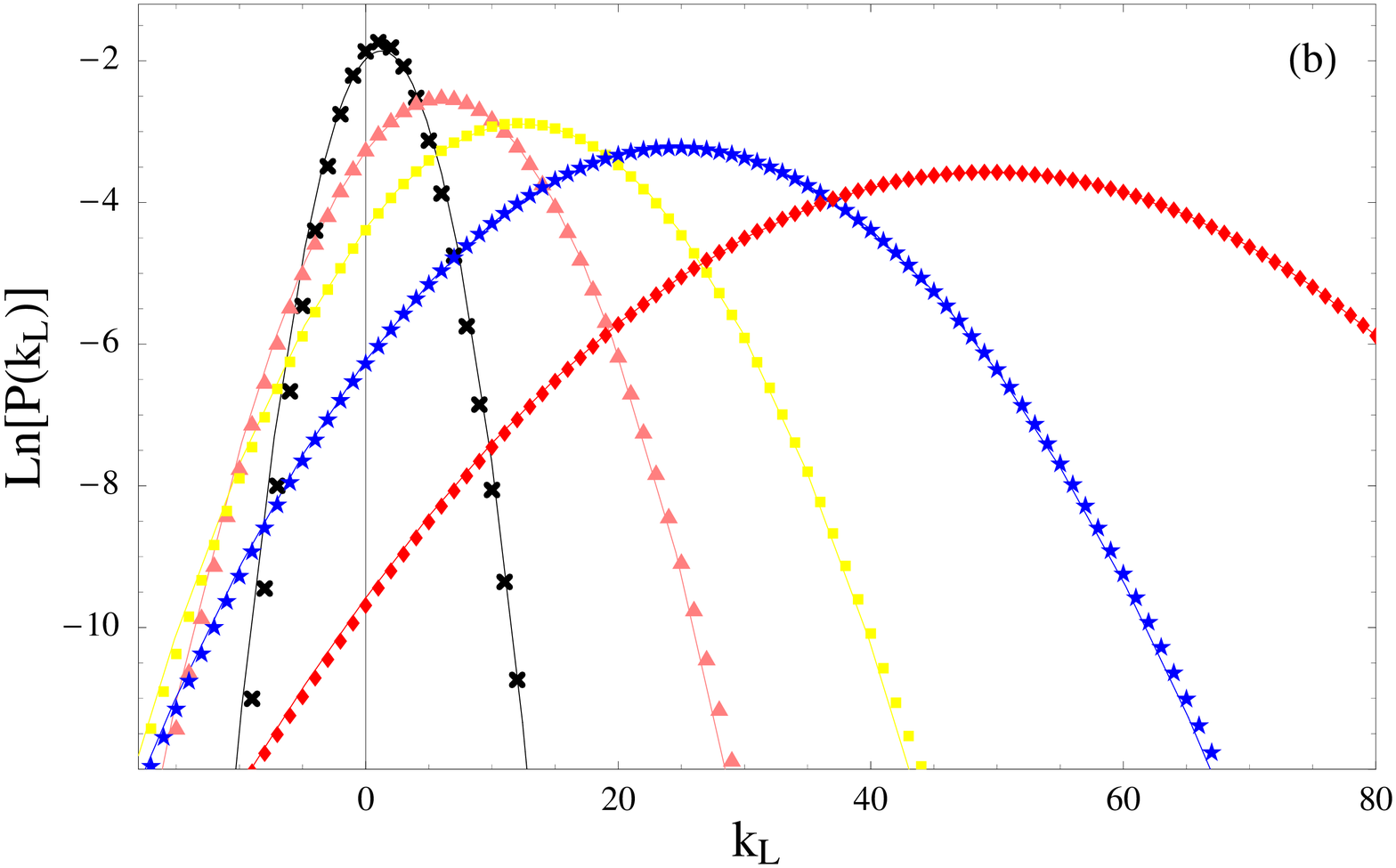}}}
\end{tabular}
\caption{a) Probability distribution of the net number of electron 
transfer $k_L$ through the left system-lead interface for different 
measurement times. 
b) The logarithm of the probability distribution highlights 
the behavior of the tails of the probability distribution.
The measurement starts when the junction is at steady state and
the different symbols correspond to different measurement times. 
The solid lines represent Gaussian fits.
$e V=0.5$ and $\beta=1$.}
\label{fig2}
\end{figure}

In Fig. \ref{fig2}a, we display the probability distribution of the net 
number of electron $k_L$ which have crossed the left lead-system interface 
for different measurement times and for a fixed temperature and bias. 
The logarithmic plot in Fig. \ref{fig2}b highlights the tails 
of the distribution. 
Positive (negative) $k_L$ represent electrons which move in 
the direction favored (unfavored) by the bias.
As the measurement time increases, the average of the probability 
distribution moves to the positive direction linearly in time at 
a speed given by the steady state current.
As expected, the longer the measurement time, the smaller the 
probability of observing a current flowing against the bias.
Since the FT applies in the long times limit, this shows that 
the FT quantifies the rare fluctuations described by the tails 
of the probability distribution.\\

\begin{figure}[p]
\centering
\begin{tabular}{c}
\vspace{0.5cm}
\rotatebox{0}{\scalebox{0.42}{\includegraphics{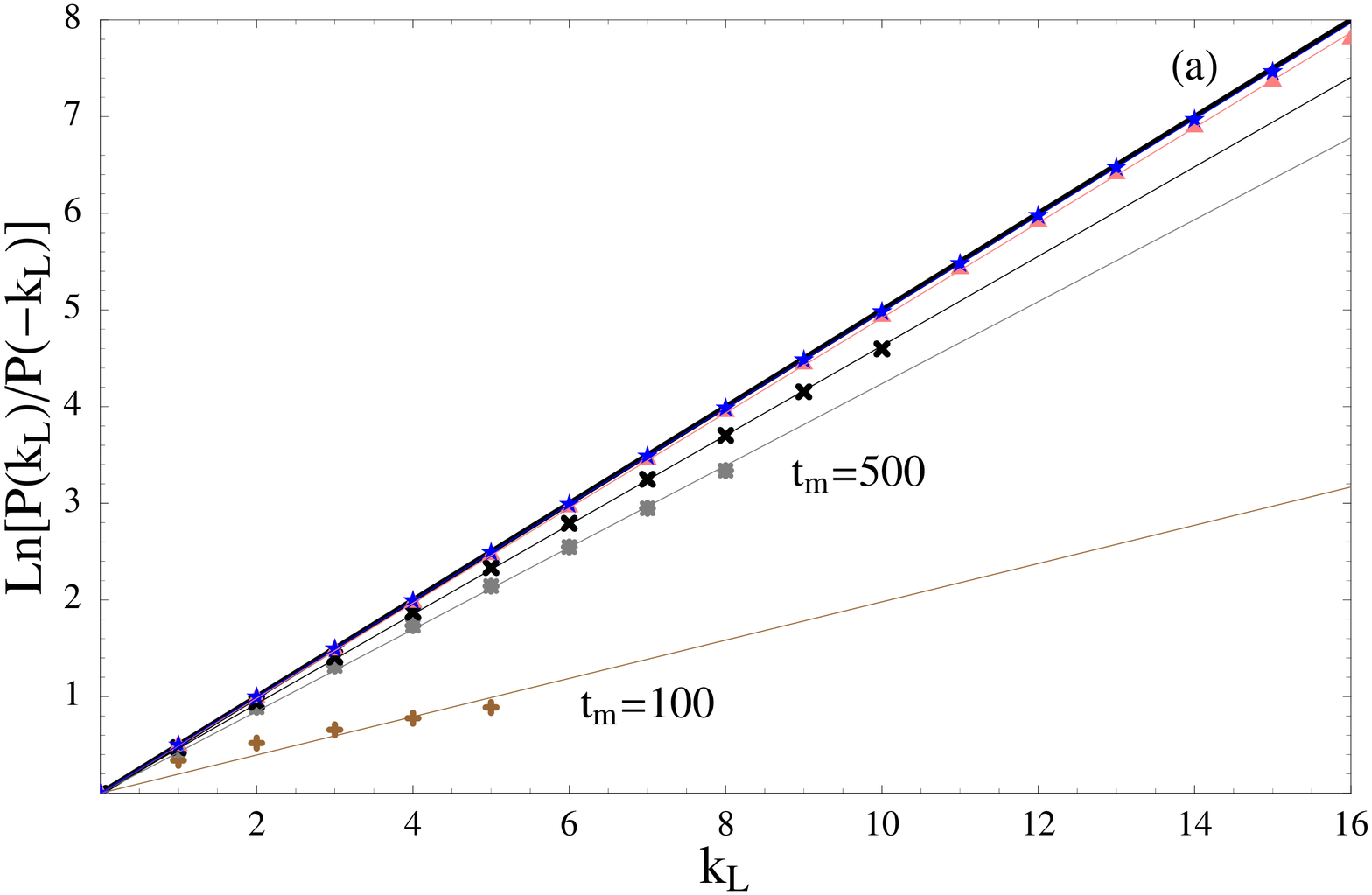}}}\\
\rotatebox{0}{\scalebox{0.42}{\includegraphics{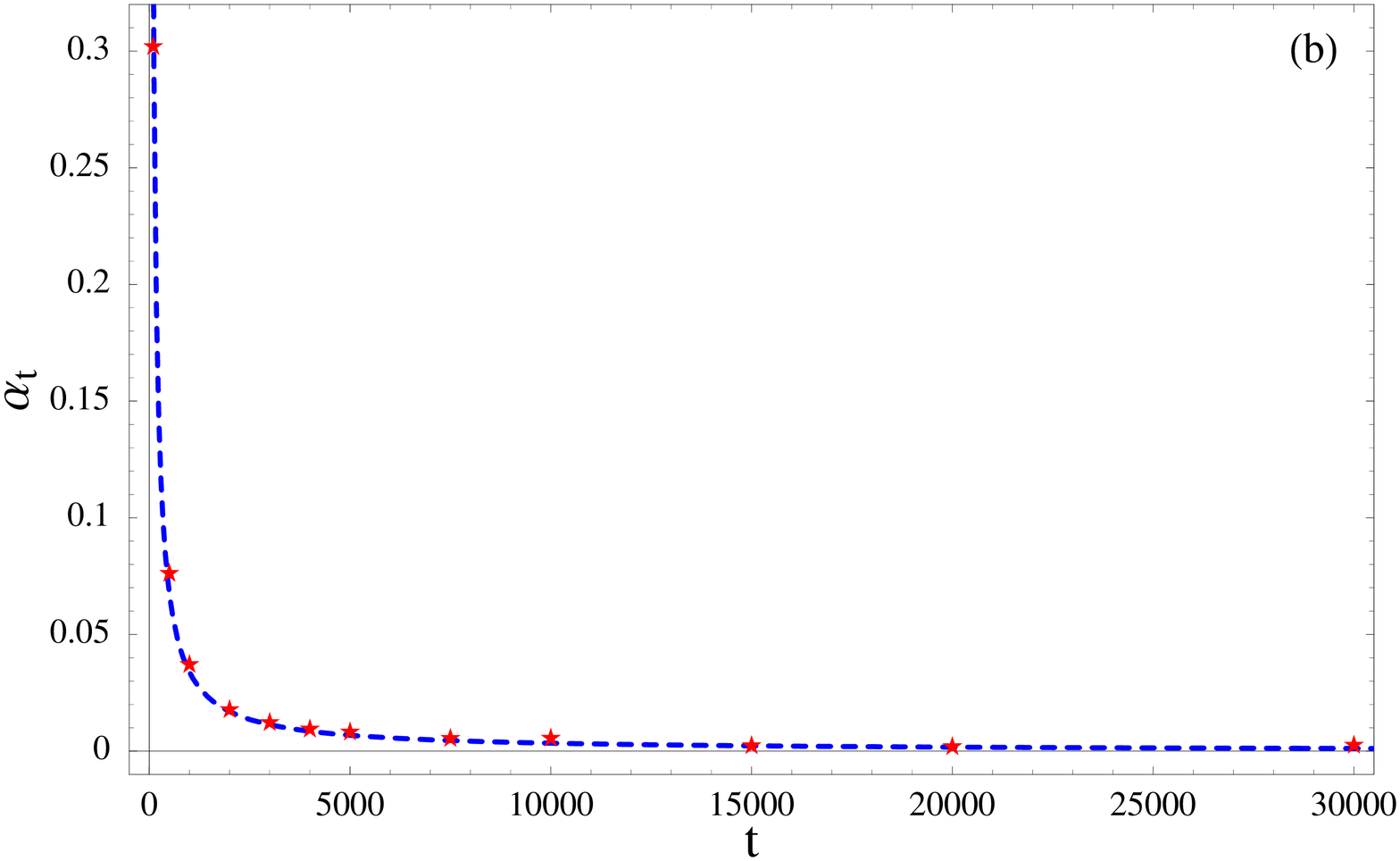}}}
\end{tabular}
\caption{
(a) The FT predicts that the logarithm of the ratio of the 
probability to measure a net number of electron transfer $k_L$ 
in the direction favored by the bias with the probability of 
measuring the opposite number $-k_L$ (which means that a net 
transfer of $k_L$ electron occurred in the direction unfavored 
by the bias) is given by $\beta e V k_L$. 
This is given by the solid black diagonal line. 
Different symbols correspond to different measurement times 
(the symbols not specifically labeled correspond to the 
same measurement times as in Fig. \ref{fig2}).
The FT requires a sufficiently large measurement time to hold.
The solid lines correspond to linear fits.
(b) The stars represent the values of $\alpha_t$ obtained from 
the linear fit of the curves from (a) and the dashed line is the 
value of $\alpha_t$ predicted by Eq.(\ref{Caaaaiaapprox3}).}
\label{fig3}
\end{figure}

Fig. \ref{fig3}a, shows the logarithm of 
$P(t,k_L)/P(t,-k_L)$ for different measurement times. 
The longer the time, the closer the results from the FT.   
The numerical results suggest that for finite times
\begin{eqnarray}
\frac{P(t,k)}{P(t,-k)} \approx e^{(\beta e V - \alpha_t) k}
\;, \label{FTaaaaahfinite}
\end{eqnarray}
where $\lim_{t \to \infty} \alpha_t=0$.
This form is not valid for very short measurement time, 
but seems to be a very good approximation for times after which 
the probability to measure at least a few electron transfers
become significant.\\

To calculate $\alpha_t$, we note that using (\ref{Caaaaea}), 
Eq. (\ref{FTaaaaahfinite}) implies that
\begin{eqnarray}
G(t,\lambda_L) \approx G(t,\beta e V - \alpha_t - \lambda_L)
\;. \label{FTaaaaahfinite2}
\end{eqnarray}
Since we do not consider very short times, (\ref{Caaaaia}) 
can be very well approximated by 
\begin{eqnarray}
\ln G(t,\lambda_L) \approx g_{m_0}(\lambda_L) t + B(\lambda_L)
\;, \label{Caaaaiaapprox}
\end{eqnarray}
where $m_0$ is the index of the eigenvalue with the smallest 
absolute value and 
\begin{eqnarray}
B(\lambda_L) &\equiv& \ln \big( \sbraket{I}{g_{m_0}(\lambda_L)} 
\sbraket{\tilde{g}_{m_0}(\lambda_L)}{\rho(0)} \big) 
\;.\label{Caaaaiaapprox1a}
\end{eqnarray}
Since $G(t,0)=G(t,\beta e V - \alpha_t)=1$, we find that
\begin{eqnarray}
g_{m_0}(\beta e V - \alpha_t) t \approx - B(\beta e V - \alpha_t)
\;. \label{Caaaaiaapprox2}
\end{eqnarray}
If we consider long enough times for which $\alpha_t \ll \beta e V$ 
and $\alpha_t/t \approx 0$, using a first [zero] order expansion of
$g_{m_0}(\beta e V - \alpha_t)$ [$B(\beta e V - \alpha_t)$] 
around $\beta e V$ and using (\ref{momentAaaaaab}), we get
\begin{eqnarray}
\alpha_t \approx \frac{B(\beta e V)}{e \mean{I_{\eta}}_{\rm st}} 
\frac{1}{t} 
\;. \label{Caaaaiaapprox3}
\end{eqnarray}
The average current can be calculated 
using $\mean{I}_{\rm st}=\sum_s^M \mean{I_{s}}_{\rm st}$
and (\ref{Caaaakc}).
Using $B(\lambda_L)=\sum_s^M B(\lambda_{(L,s)})$,
where $B(\lambda_{(L,s)})=\ln \big( \sbraket{I}{g_{-}(\lambda_{(L,s)})} 
\sbraket{\tilde{g}_{-}(\lambda_{L,s})}{\rho(0)} \big)$ and
$\rho(0)$ correspond to steady state, we find that
\begin{eqnarray}
B(\beta e V)
= \sum_s^M \ln \frac{( v_{ss}^{(L)}+e^{-\beta e V} w_{ss}^{(L)} )
\big( e^{\beta e V} w_{ss}^{(L)} (w_{ss}^{(L)}+w_{ss}^{(R)})^2 
+ v_{ss}^{(L)} (w_{ss}^{(L)}+e^{\beta e V} w_{ss}^{(R)})^2 \big)}
{\big(e^{\beta e V} v_{ss}^{(L)} w_{ss}^{(R)} +w_{ss}^{(L)} 
(v_{ss}^{(L)}+w_{ss}^{(L)}+w_{ss}^{(R)})\big)^2}
\;.\label{Caaaaiaapprox1b}
\end{eqnarray}
Notice that since $\exp{\{B(\lambda_{(L,s)})\}} \ge 1$, then
$\alpha_t \ge 0$ (the equality only holds when $\beta e V=0$).\\

\begin{figure}[p]
\centering
\rotatebox{0}{\scalebox{0.42}{\includegraphics{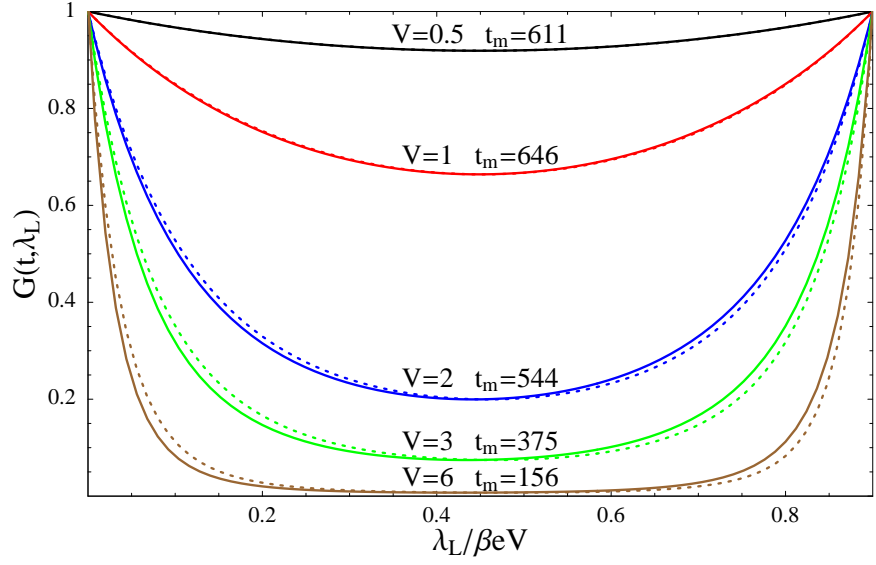}}}
\caption{
Generating function calculated for different values of the bias 
and at measurement times $t_m$ such that $\alpha_{t_m}=0.1 \beta e V$. 
$t_m$ is calculated by solving $G(t_m,0.9 \beta e V)=1$.
The solid line is $G(t_m,\lambda_L)$ and the dotted line is 
$G(t_m,0.9 \beta e V-\lambda_L)$. 
The difference between the two curves is a measure of the 
breakdown of the symmetry (\ref{FTaaaaahfinite2}).}
\label{fig3bis}
\end{figure}

Figure \ref{fig3}b shows that our estimate for $\alpha_t$ is in 
excellent agreement with the values obtained by linearly fitting 
the results of Fig. \ref{fig3}a.
It should be noted that the results presented on Fig. \ref{fig2}-\ref{fig3} 
correspond to small bias. 
For larger bias the probability of the backward processes $P(-k)$ becomes 
very small which limits the numerical accuracy.
However, the GF is still numerically accessible for high bias.
Fig. \ref{fig3bis} shows that the GF symmetry (\ref{FTaaaaahfinite2}) on 
which our methods relies is not perfectly preserved for larger bias. 
It is therefore expected that the accuracy of our method 
decreases with increasing bias.\\ 

One can see on Fig. \ref{fig2} that the probability distribution
can be reasonably well fitted by a Gaussian.
Deviations can be observed at very short times or for the tails 
of the distributions. 
The GF of a Gaussian probability distribution
$P(t,k)=(2 \pi \sigma^2_t)^{-1/2}\exp{\{-(k-\mean{k}_t)^2/
(2 \sigma^2_t)\}}$ is given by 
$G(t,\lambda)=\exp{\{\lambda^2 \sigma^2_t /2-\lambda \mean{k}_t\}}$.
The nonzero solution of $G(t,\lambda)=1$ is 
$\lambda_0=2 \mean{k}_t/\sigma^2_t=2/F(t)$, where $F(t)$ is the 
time dependent Fano parameter associated with the net number of 
charge transferred through an interface. 
The GF has the symmetry $G(t,\lambda)=G(t,\lambda_0-\lambda)$.
Using the results of section \ref{AverageGF} to calculate 
$F(\infty)$, we find that for very long measurement times
$\lambda_0=\beta e V + {\cal O}(V^3)$.
This indicates that it is only in the low bias limit that 
the Gaussian approximation can be trusted for describing the 
tails of the probability distribution which characterize the FT.
This was confirmed by numerically studying the GF.
We finally notice that our estimate of $\alpha_t$ is 
exact when the Gaussian approximation is satisfied, 
but can also be applied to non-Gaussian distributions as long 
as (\ref{FTaaaaahfinite}) or (\ref{FTaaaaahfinite2}) remains 
a good assumption.\\

\begin{figure}[p]
\centering
\begin{tabular}{c@{\hspace{0.5cm}}c}
\rotatebox{0}{\scalebox{0.3}{\includegraphics{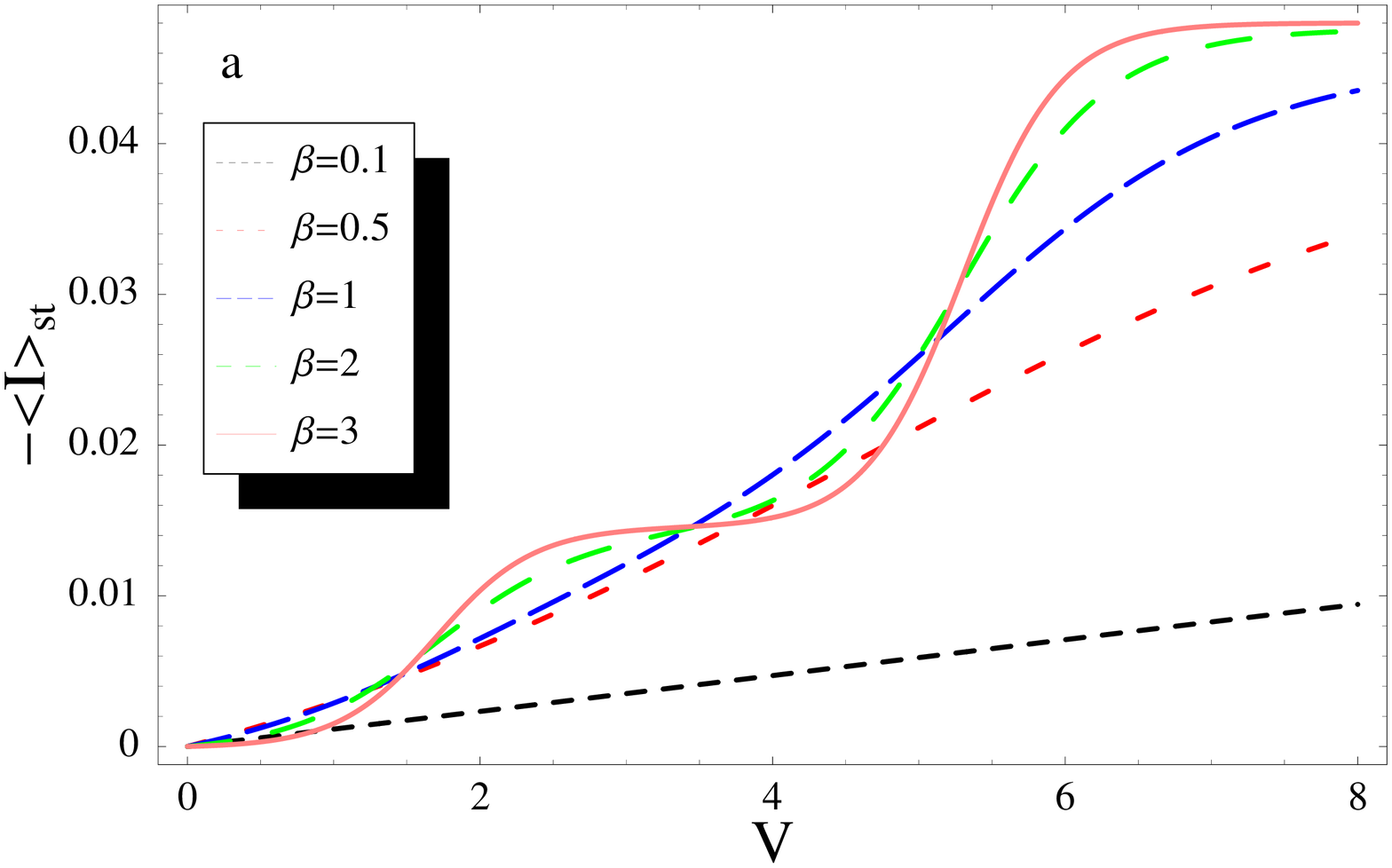}}} &
\rotatebox{0}{\scalebox{0.3}{\includegraphics{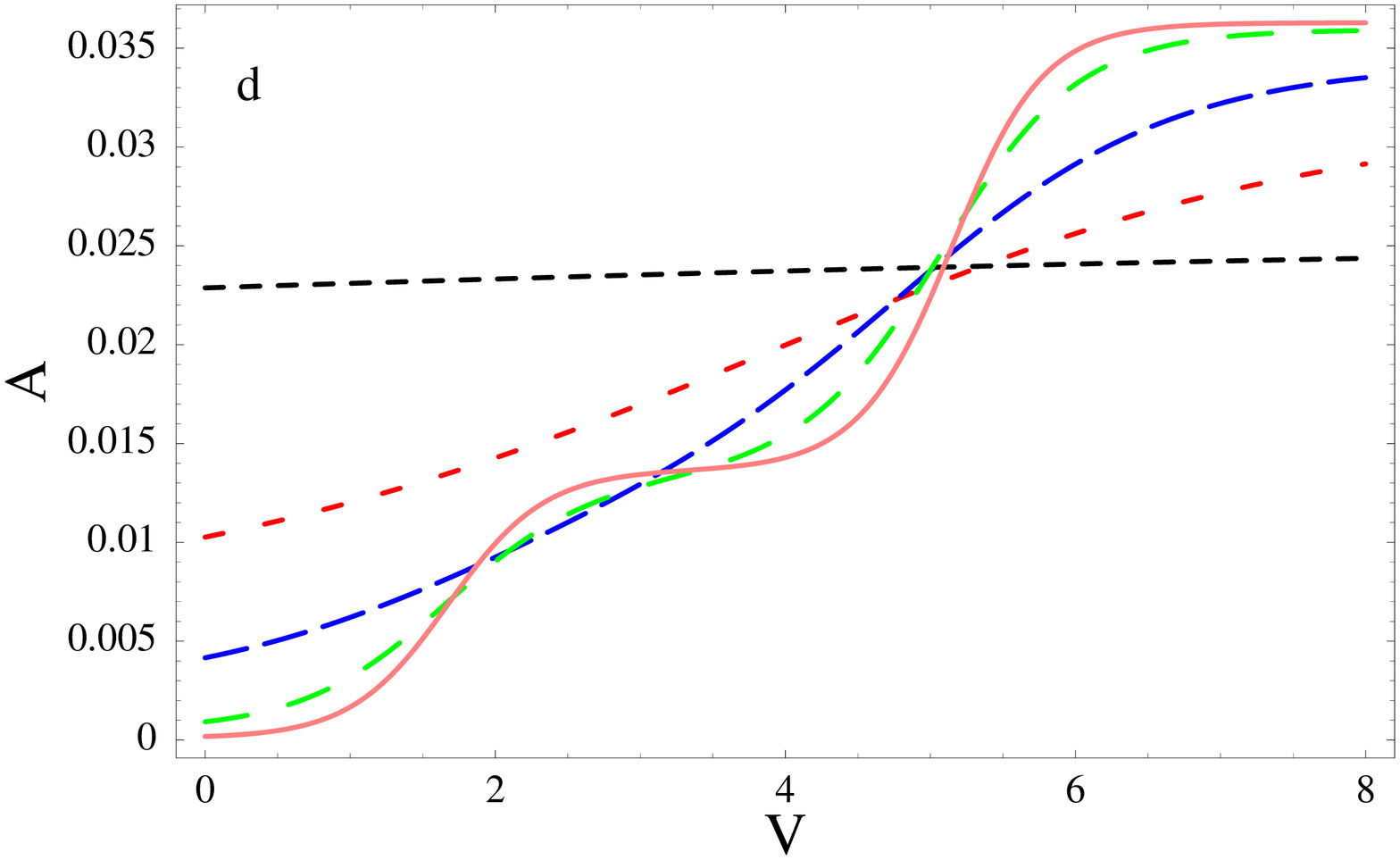}}} \\
\rotatebox{0}{\scalebox{0.3}{\includegraphics{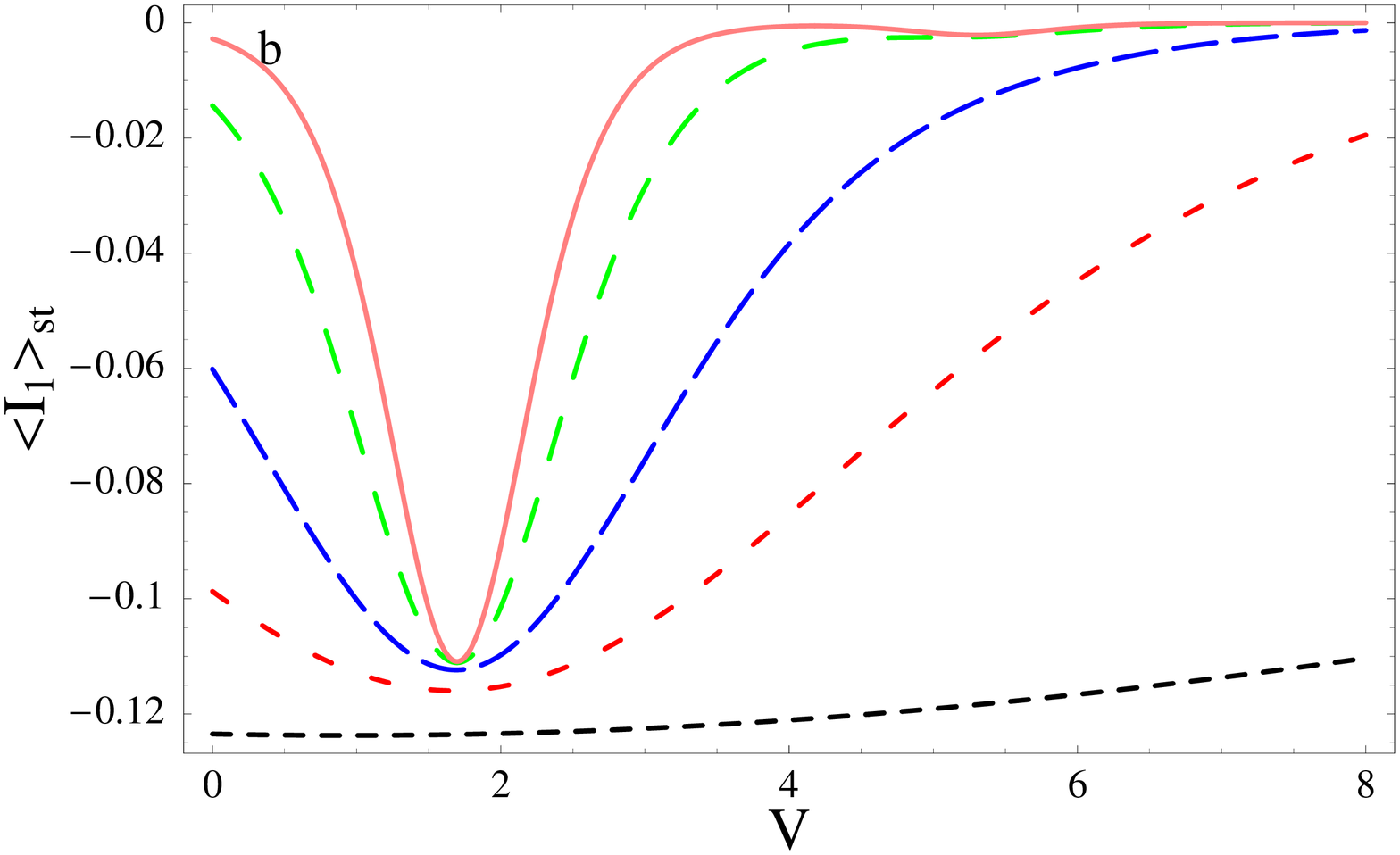}}} &
\rotatebox{0}{\scalebox{0.3}{\includegraphics{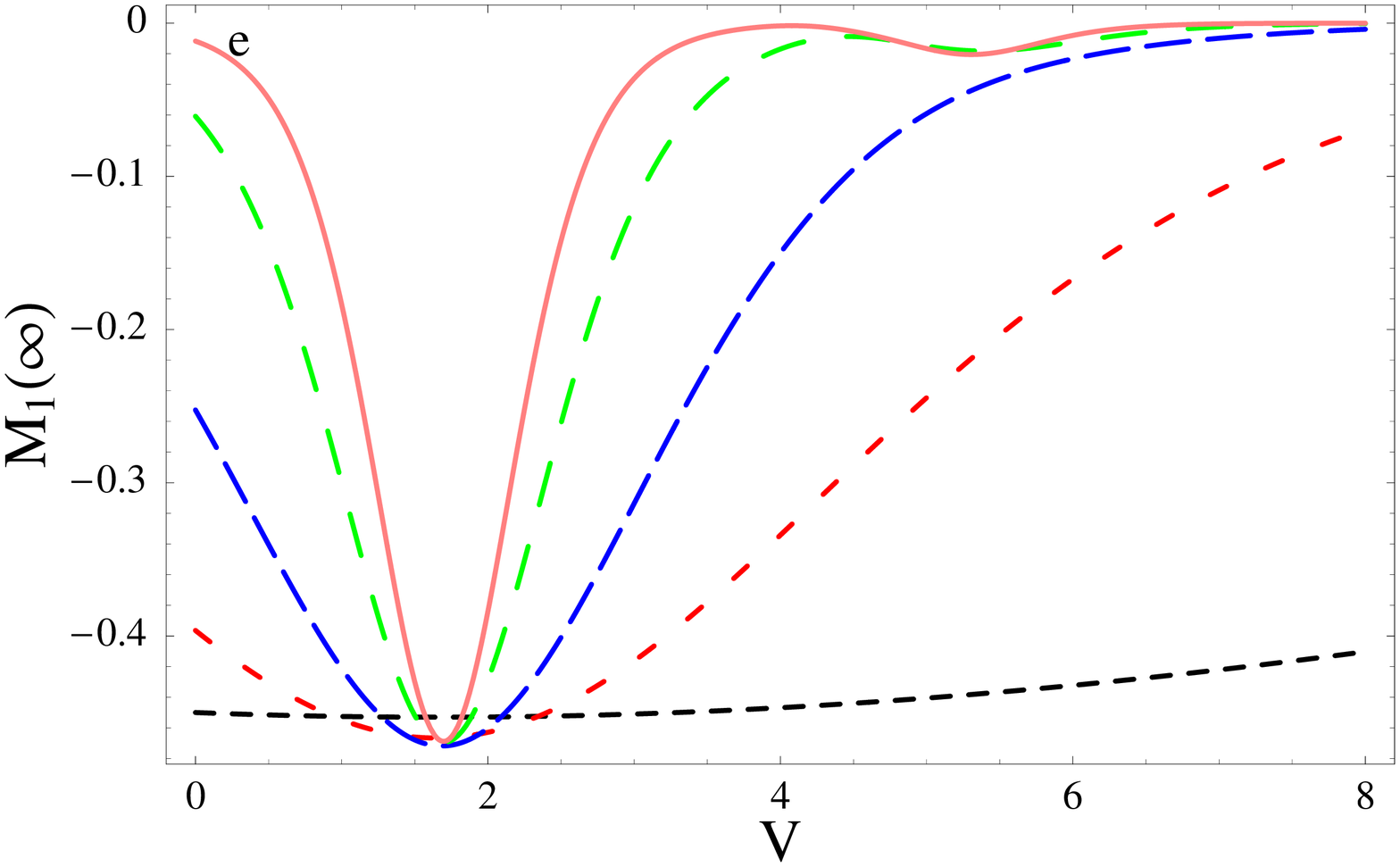}}} \\
\rotatebox{0}{\scalebox{0.3}{\includegraphics{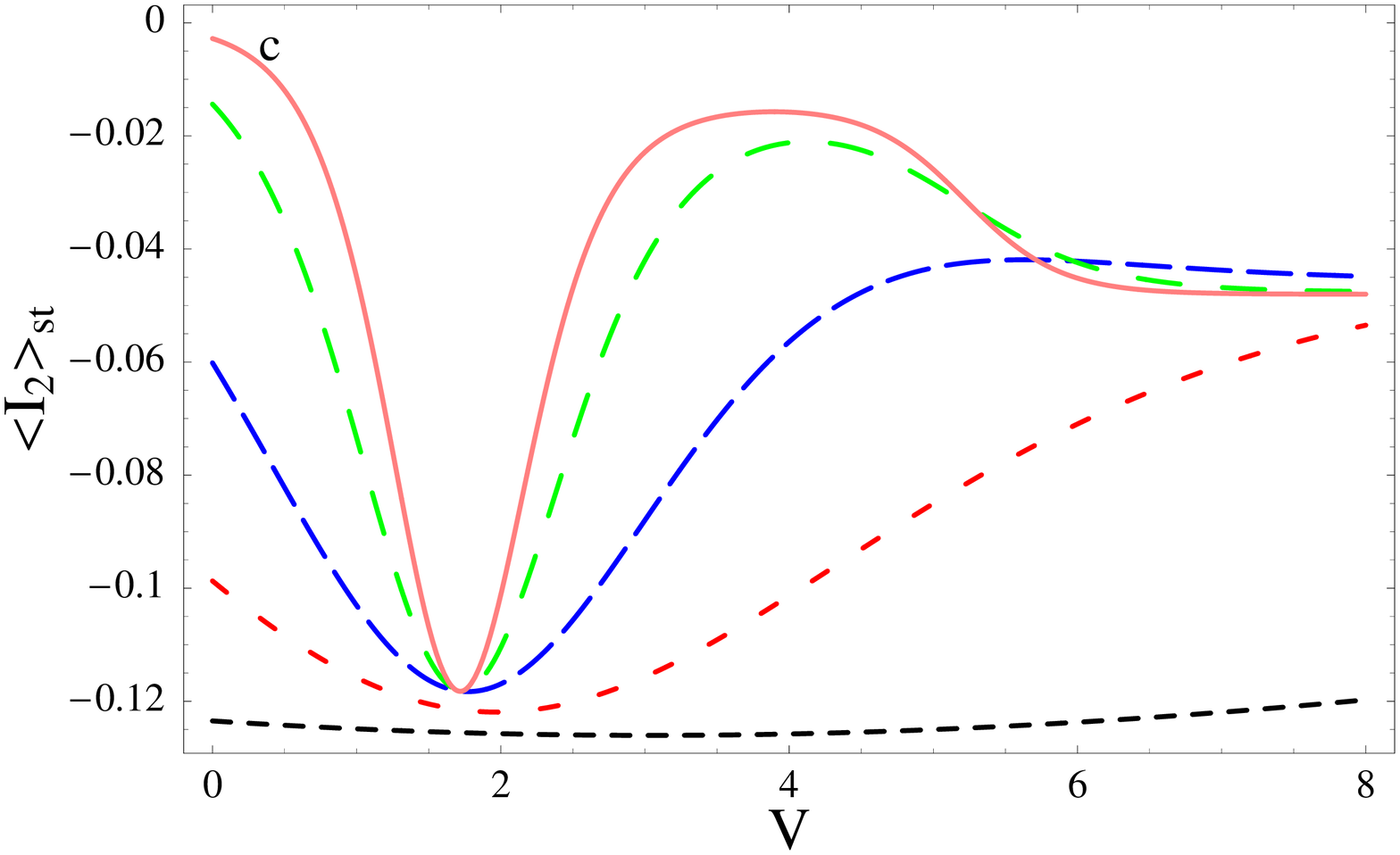}}} &
\rotatebox{0}{\scalebox{0.3}{\includegraphics{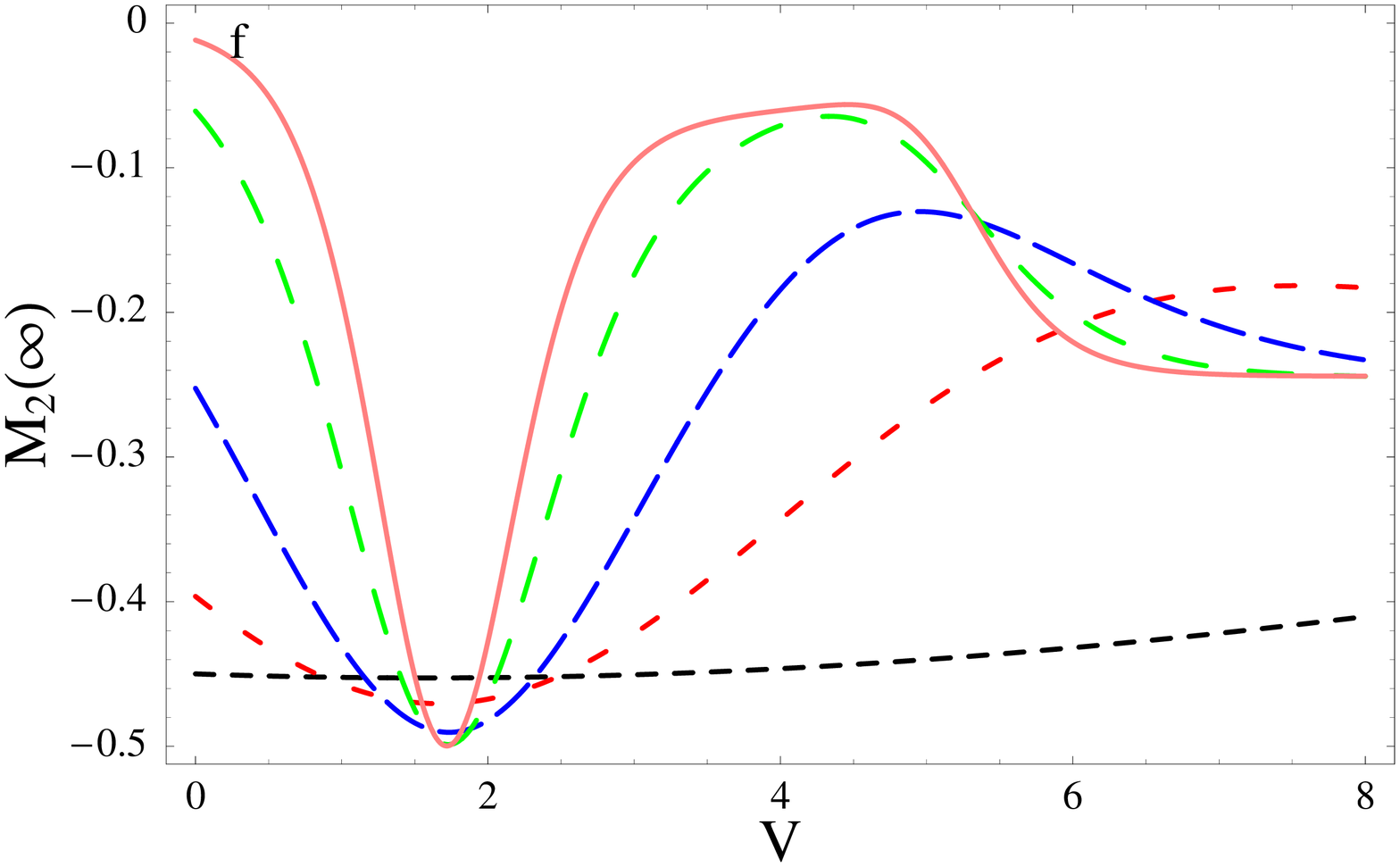}}} 
\end{tabular}
\caption{
a) Average steady state net current 
$\langle I \rangle_{\rm st}$.  
The bias is at resonance with the dot levels for 
$eV=\epsilon_1=1.697$ and $eV=\epsilon_2=5.303$.
b) Average current due to the electron exiting the dots 
from the left side of the junction (processes 
$\eta=1$ on Fig. \ref{fig1}).
c) Average current due to the electron entering the dots 
from the left side of the junction 
(processes $\eta=2$ on Fig. \ref{fig1}). 
d) Zero frequency power spectrum of the steady state 
net current [$A_1+A_2$ using (\ref{Caaaakb})].  
e) Mandel parameter associated to the process $\eta=1$. 
f) Mandel parameter associated to the process $\eta=2$.
All quantities plotted in this figure are functions 
of the bias and are represented for five different temperature.}
\label{fig4}
\end{figure}

The average steady-state current associated with the 
net number of electrons crossing a given junction interface is 
plotted as a function of the bias in Fig. \ref{fig4}a. 
This typical current-voltage characteristic 
shows that the current increases by steps each time the bias is 
large enough to make a new orbital contribute to the current
(when $e V = \epsilon_1=1.697$ or $e V = \epsilon_2=5.303$). 
The current associated with the process $\eta=1$ and $\eta=2$ is 
plotted on Fig. \ref{fig4}b and \ref{fig4}c.
The current on Fig. \ref{fig4}a is given by the difference 
between Fig. \ref{fig4}c and Fig. \ref{fig4}b.
Temperature, when increased, has the effect of smoothing these 
steps and reducing the current because thermal fluctuations 
tend to equalize the forward and backward currents. 
Ohm's law is recovered for 
$\beta (\epsilon_2-\epsilon_1) < 1$.
The zero frequency power spectrum associated to the total net 
current through a system interface is plotted as a function of 
the bias in Fig. \ref{fig4}d.
The Mandel parameter associated with the $\eta=1$ and 
$\eta=2$ processes is plotted in Fig. \ref{fig4}e and \ref{fig4}f. 
We see that the deviations from Poisson statistics are always 
strong and tend to vanish only when the currents associated 
with $\eta=1$ and $\eta=2$ vanish.
This indicates that the various type of electron 
transfer are highly correlated with each other. 
The Mandel parameter is always negative, indicating 
subpoissonian (antibunching) statistics. 
This has been experimentally observed in \cite{Hirayama}.

%%%%%%%%%%%%%%%%%%%%%%%%%%%%%%%%%%%%%%%%%%%%%%%%%%%%%%%%%%%%%%%%%%%%%%%%%%%%%%%%%%%%%%%%%%%%%%%%%%%%%%%%%%
\section{Discussion}\label{conclusion}

Many different types of fluctuation theorems (FT) have 
been derived for stochastic dynamical systems. 
These differ by the mechanism used to drive 
the system out of equilibrium.
In the first case \cite{Crooks99,Crooks00,Seifert}, the system is 
closed and driven by a time-dependent force which makes the rate 
matrix of the birth and death master equation time dependent. 
When the driving stops, the system will eventually reach 
equilibrium because the rate matrix is detailed balanced \cite{Crooks99}.
In the second case \cite{Lebowitz,Gaspard1,Seifert,Maes2}, the 
system is open and the the rate matrix is not detailed balanced.
Even without driving, the system will eventually reach a 
nonequilibrium steady state.
A third class of FT \cite{Hatano99,Hatano01,Bustamante04}
considers the fluctuation of an entropy associated 
with the excess heat produced when a time dependent driving 
induces transitions between different nonequilibrium steady states.
This paper focuses on the second case.\\

So far, we have assumed that the two leads of the junction 
have the same temperature but different chemical potentials 
$\mu_{L} = \mu_0 + \Delta \mu$ and $\mu_{R} = \mu_0$.
One could wonder what happens to the FT if one considers 
different temperatures for the two leads
$\beta_{L} = \beta_0 + \Delta \beta = $ and $\beta_{R}= \beta_0$.
In such case, the argument of the exponential on the r.h.s. of 
(\ref{QMEaaaaad}) becomes $x \equiv \beta_0 \Delta \mu 
- \Delta \beta (\epsilon_s-\mu_0) + \Delta \beta \Delta \mu $. 
This implies for the orbital GF that the analogue of the symmetry 
(\ref{Caaaai}) becomes $S_s(\lambda_{(y,s)})=S_s(x-\lambda_{(y,s)})$.
However, since $x$ is different for each orbital (due to $\epsilon_s$),
the many body GF, which is given by the sum of the orbital GFs, does not
possess the analogue of the symmetry (\ref{Caaaaib}).
It is therefore only for a single orbital model that a FT 
$P(t,k)/P(t,-k) = \exp{\{x k\}}$ holds.\\

In summary, we have applied the quantum master equation derived in 
\cite{HarbolaEsposito} for calculating the counting statistics of 
electrons tunneling through a quantum junction made of a system 
embedded between two leads.
Using a generating function formalism, we derived an evolution 
equation for the generating operator which allows to calculate the 
time dependent probability distribution of electron transfer events.
This equation can be solved analytically because the many-body
generating operator is a product of single orbital generating operators.
We then demonstrated that the net number of electrons crossing a given 
system-lead interface satisfies a FT at long times. 
This implies that measuring a net number of electron transfer in the 
direction favored by the bias is exponentially more probable 
than measuring it in the opposite direction.
Since the argument in the exponential is the work needed to transfer 
the measured electrons through the junction, this fluctuation 
theorem can be viewed as a Crooks relation \cite{Crooks99,Crooks00}.
We furthermore described how the moments of the current distribution can 
be deduced from the electron counting statistics and gave analytical 
expressions for currents and power spectra.
Numerically calculations of the probability distribution for the 
electron counts for a model of two coupled quantum dots demonstrated 
that the FT becomes valid for long measurement times. 
A method to calculate the finite-time deviations was proposed.\\

Several future extensions of this work are called for.
The first one is to find if electron-electron interactions affect the FT. 
Another one is to investigate if the FT still holds in a system where 
the population dynamics couple to the coherence dynamics (e.g. a 
quantum junction externally driven by a laser).
In analogy to the excess heat \cite{Hatano01}, one could also consider 
fluctuations of excess currents, produced during transition between 
steady-states. 

%%%%%%%%%%%%%%%%%%%%%%%%%%%%%%%%%%%%%%%%%%%%%%%%%%%%%%%%%%%%%%%%%%%%%%%%%%%%%%%%%%%%%%%%%%%%%%%%%%%%%%%%%%

\section*{Acknowledgment}

The support of the National Science Foundation (Grant No. CHE-0446555) 
and NIRT (Grant No. EEC 0303389) is gratefully acknowledged.
M. E. is partially supported by the FNRS Belgium
(collaborateur scientifique).\\

%%%%%%%%%%%%%%%%%%%%%%%%%%%%%%%%%%%%%%%%%%%%%%%%%%%%%%%%%%%%%%%%%%%%%%%%%%%%%%%%%%%%%%%%%%%%%%%%%%%%%%%%%%

\appendix
%%%%%%%%%%%%%%%%%%%%%%%%%%%%%%%%%%%%%%%%%%%%%%%%%%%%%%%%%%%%%%%%%%%%%%%%%%%%%%%%%%%%%%%%%%%%%%%%%%

%%%%%%%%%%%%%%%%%%%%%%%%%%%%%%%%%%%%%%%%%%%%%%%%%%%%%%%%%%%%%%%%%%%%%%%%%%%%%%%%%%%%%%%%%%%%%%%%%%
\section{Trajectory picture for the QME dynamics} \label{trajectories}

Using the interaction representation, we can recast Eq. (\ref{ecountAaaaa}) as
\begin{eqnarray}
\sket{\dot{\rho}_{I}(t)} = \hat{\Gamma}(t) \sket{\rho_{I}(t)}  \;,
\label{ecountAaaak}
\end{eqnarray}
where
\begin{eqnarray}
\sket{\rho_{I}(t)} \equiv \hat{{\cal U}}_{0}(0,t) \sket{\rho(t)}
\label{ecountAaaaj}
\end{eqnarray}
and
\begin{eqnarray}
\hat{\Gamma}(t) = \hat{{\cal U}}_{0}(0,t) \hat{\Gamma} \hat{{\cal U}}_{0}(t,0) \;,
\label{ecountAaaal}
\end{eqnarray}
where $\hat{{\cal U}}_{0}(0,t) \equiv \exp{[-\hat{{\cal M}}_{0} t]}$.
$\hat{{\cal M}}_{0}=\hat{{\cal L}} + \hat{\gamma}$ describes 
the system dynamics in absence of electron transfer.
The formal solution of (\ref{ecountAaaak}) reads
\begin{eqnarray}
\sket{\rho_{I}(t)} &=&
\exp_{+}{[ \int_{0}^{t} d\tau \hat{{\Gamma}}(\tau) ]} \sket{\rho_{I}(0)}
= \sum_{k=0}^{\infty} \sket{\rho^{(k)}_{I}(t)} \;,
\label{ecountAaaam}
\end{eqnarray}
where
\begin{eqnarray}
\sket{\rho^{(k)}_{I}(t)} = \int_{0}^{t} d\tau_{k} \int_{0}^{\tau_{k}}
d\tau_{k-1} \ldots \int_{0}^{ \tau_{2}} d \tau_{1} \hat{\Gamma}(\tau_{k})
\hat{\Gamma}(\tau_{k-1}) \ldots \hat{\Gamma}(\tau_{1}) \sket{\rho_{I}(0)}
= \int_{0}^{t} d\tau \hat{\Gamma}(\tau) \sket{\rho^{(k-1)}_{I}(\tau)}
\;. \label{ecountAaaao}
\end{eqnarray}
Multiplying Eq. (\ref{ecountAaaao}) by $\hat{{\cal U}}_{0}(t,0)$,
we get
\begin{eqnarray}
\sket{\rho^{(k)}(t)} =\int_{0}^{t} d\tau_{k} \int_{0}^{\tau_{k}} d\tau_{k-1}
\ldots \int_{0}^{ \tau_{2}} d \tau_{1}
\hat{{\cal U}}_{0}(t,\tau_{k}) \hat{\Gamma}
\hat{{\cal U}}_{0}(\tau_{k},\tau_{k-1}) \hat{\Gamma}
\ldots \hat{{\cal U}}_{0}(\tau_{2},\tau_{1}) \hat{\Gamma} \hat{{\cal U}}_{0}(\tau_{1},0)
\sket{\rho(0)} \;. \label{ecountAaaaob}
\end{eqnarray}
This is the density matrix conditional to the transfer of $k$
electrons between the system and the leads, irrespective of the type
of transfer $\eta$ or orbital $s$.\\

We shall denote the number of electron transfers of type 
$\eta$ through the $s$ orbital by $k_{\nu}$ where $\nu=(\eta,s)$.
The number of electron transfers of type $\eta$ disregarding the orbital 
is $k_{\eta} = \sum_{s=1}^{M} k_{(\eta,s)}$ and the number of electron transfer 
through the orbital $s$ disregarding the type of transfer 
is $k_{s} = \sum_{\eta=1}^{4} k_{(\eta,s)}$.
We have that $k = \sum_{\nu} k_{\nu}$.
We define
\begin{eqnarray}
&&\bold{k} = k_{(1,1)},k_{(1,2)},\cdots,k_{(1,M)},
k_{(2,1)},k_{(2,2)},\cdots,k_{(2,M)},\cdots,k_{(4,1)},k_{(4,2)},\cdots,k_{(4,M)}
\label{ecountAaaaob1}\\
&&\bold{k}_{s} = k_{(1,s)},k_{(2,s)},k_{(3,s)},k_{(4,s)} \nonumber\\
&&\bold{k}_{\eta} = k_{(\eta,1)},k_{(\eta,2)},\cdots,k_{(\eta,M)} \;.
\nonumber
\end{eqnarray}
%so that we can use the compact notation
%\begin{eqnarray}
%&&\underline{\vec{k}} \cdot \underline{\vec{k}}' =
%\sum_{\eta=1}^{4} \underline{k}_{\eta} \cdot \underline{k}_{\eta}'
%\ \ {\rm where}  \  \
%\underline{k}_{\eta} \cdot \underline{k}_{\eta}' =
%\sum_{s=1}^{M} k_{(\eta,s)} k_{(\eta,s)}' \label{ecountAaaaob5}\\
%&&\sum_{\underline{\vec{k}}} = \sum_{\eta=1}^{4}
%\sum_{\underline{k}_{\eta}} = \sum_{\eta=1}^{4} \sum_{s=1}^{M}
%\sum_{k_{(\eta,s)}=0}^{\infty} \label{ecountAaaaob6} \;.
%\end{eqnarray}
A trajectory $(\nu,t)_{(\tau)}$ records (from left to right) the 
sequence of electron transfer events during a time $\tau$, by labeling 
them according to the type of transfer, the relevant orbital and
the time at which a given transfer occurs
$(\nu,t)_{(\tau)}=(\nu_1,t_1),(\nu_2,t_2), \cdots
,(\nu_k,t_k)$. A sequence $(\nu)_{(\tau)}$ is a trajectory
where the transfer time is not recorded
$(\nu)_{(\tau)}=\nu_1,\nu_2, \cdots ,\nu_k$.\\
We can now decompose $\rho_{I}(t)$ as
\begin{eqnarray}
\sket{\rho_{I}(t)}= \sum_{\bold{k}} \sket{\rho_{I}^{(\bold{k})}(t)} \;,
\label{ecountAaaan}
\end{eqnarray}
where the sum is over all the component $k_{\nu}$ of $\bold{k}$ and 
runs from zero to infnity.
Using (\ref{ecountAaaaqamain}), the probability to measure $\bold{k}$ 
electrons transferred to the leads at time $t$ is given by
\begin{eqnarray}
P(t,\bold{k})
&=& \sbra{I} \hat{{\cal U}}_{0}(t,0) \sket{\rho_{I}^{(\bold{k})}(t)} \;.
\label{ecountAaaaqa}
\end{eqnarray}
The master equation (\ref{ecountAaaaa}) preserves the trace so that 
the normalization condition $\sum_{\bold{k}} P(t,\bold{k}) = 1$ is satisfied.\\

We next define the elementary probability density of a given trajectory
$(\nu,\tau)_{(t)}$ which contains $k$ electron transfer events at time
$\tau_1,\ldots,\tau_{k}$ during a time interval $t$ as
\begin{eqnarray}
\Pi[(\nu,\tau)_{(t)}] \equiv \sbra{\trace} \hat{{\cal U}}_{0}(t,0)
\hat{\Gamma}_{\nu_k}(\tau_{k})
\hat{\Gamma}_{\nu_{k-1}}(\tau_{k-1})
\ldots \hat{\Gamma}_{\nu_1}(\tau_{1}) \sket{\rho_{I}(0)} \;.
\label{ecountAaaaoc}
\end{eqnarray}
The probability of a sequence $(\nu)_{(t)}$ with $k$ events
is obtained by time integrating (\ref{ecountAaaaoc})
\begin{eqnarray}
\Pi[(\nu)_{(t)}] &=& \int_{0}^{t} d\tau_{k} \int_{0}^{\tau_{k}}
d\tau_{k-1} \ldots \int_{0}^{ \tau_{2}} d \tau_{1} \Pi[(\nu,\tau)_{(t)}] \;.
\label{ecountAaaaod}
\end{eqnarray}
Using Eq. (\ref{ecountAaaaob}), and since our equation conserves the trace,
we see that $\sum_{(\nu)_{(t)}} \Pi[(\nu)_{(t)}] = 1$,
where $\sum_{(\nu)_{(t)}}$ is the sum over all possible electron transfer
sequences (without keeping track of transfer times).
The trace of (\ref{ecountAaaao}) can now be written using (\ref{ecountAaaaoc})
and (\ref{ecountAaaaod}) as
\begin{eqnarray}
P(t,\bold{k}) = \sum_{(\nu)_{(t)} \in \bold{k}} \Pi[(\nu)_{(t)}] \;,
\label{ecountAaaaoe}
\end{eqnarray}
where the summation is restricted to sequences such that the number of transfer
events is $\bold{k}$.\\
Because a number $\bold{k}$ of electrons at time $t$ can be realized by the four 
types of electron transfer processes (Fig. \ref{fig1}) and via $M$ different 
orbitals, we have
\begin{eqnarray}
\sket{\rho_{I}^{(\bold{k})}(t)} &=& \sum_{\nu}
\int_{0}^{t} d\tau \hat{\Gamma}_{\nu}(\tau)
\sket{\rho_{I}^{(\bold{k}-\bold{1}_{\nu})}(\tau)} \;,
\label{ecountAaaap}
\end{eqnarray}
where $\bold{1}_{\nu} = 0,\cdots,0,1,0,\cdots,0$, 
where $1$ is at the position $\nu$ in the sequence. \\
Using the interaction picture of the GO, 
we can rewrite (\ref{ecountAaaap}) as
\begin{eqnarray}
\sket{G_I(t,\boldsymbol{\lambda})} = 
\int_{0}^{t} d\tau \big( \sum_{\nu} e^{\lambda_{\nu}}
\hat{\Gamma}_{\nu}(\tau) \big)
\sket{G_I(\tau,\boldsymbol{\lambda})} \;.
\label{ecountAaaas}
\end{eqnarray}
By taking the time derivative and going back to the Schr\"odinger 
picture, we get Eq. (\ref{ecountAaaatab}).

%%%%%%%%%%%%%%%%%%%%%%%%%%%%%%%%%%%%%%%%%%%%%%%%%%%%%%%%%%%%%%%%%%%%%%%%%%%%%%%%%%%%%%%%%%%%%%%%%%
\section{Solution of the GF} \label{singlebody}

The generator of (\ref{ecountAaaatab}) can be written as a 
sum of contributions of each orbital
\begin{eqnarray}
\hat{{\cal W}}(\boldsymbol{\lambda}) = \sum_{s=1}^{M} 
\hat{{\cal W}}_s(\boldsymbol{\lambda}_{s})
\;, \label{singleAaaa1}
\end{eqnarray}
where
\begin{eqnarray}
\hat{{\cal W}}_s(\boldsymbol{\lambda}_{s})
= \hat{{\cal L}}_s + \hat{\gamma}_{s} 
+ \sum_{\eta=1}^{4} e^{\lambda_{(\eta,s)}} \hat{\Gamma}_{(\eta,s)} 
\label{singleAaaa2}
\end{eqnarray}
and $\hat{{\cal L}}_s \equiv - i \epsilon_s [ c_s^{\dagger} c_s, \cdot ]$.
The GO can therefore be factorized as a tensor product of 
orbital GO 
\begin{eqnarray}
\sket{G(t,\boldsymbol{\lambda})} = 
\prod_{s=1}^{M} \sket{G_s(t,\boldsymbol{\lambda}_{s})}\otimes \;,
\label{singleAaaa3}
\end{eqnarray}
which evolve independently according to
\begin{eqnarray}
\sket{\dot{G}_s(t,\boldsymbol{\lambda}_{s})}= 
\hat{{\cal W}}_s(\boldsymbol{\lambda}_{s}) \sket{G_s(t,\boldsymbol{\lambda}_{s})} \;.
\label{singleAaaa3}
\end{eqnarray}
Projecting Eq. (\ref{singleAaaa3}) into the system eigenbasis 
and using the notation
\begin{eqnarray}
\sbraket{n_1 \cdots n_M; n_1' \cdots n_M'}{G(t,\boldsymbol{\lambda})}
%=\bra{n_1 \cdots n_M} G(t,\boldsymbol{\lambda}) \ket{n_1 \cdots n_M}
=\prod_{s=1}^{M} \sbraket{n_s;n_s'}{G_s(t,\boldsymbol{\lambda}_{s})}
=\prod_{s=1}^{M} G_{n_s;n_s'}(t,\boldsymbol{\lambda}_{s}) \;,
\label{singleAaaa4}
\end{eqnarray}
we get
\begin{eqnarray}
\left(
\begin{array}{c}
\dot{G}_{1;1}(t,\boldsymbol{\lambda}_{s}) \\ \dot{G}_{0;0}(t,\boldsymbol{\lambda}_{s}) \\
\dot{G}_{1;0}(t,\boldsymbol{\lambda}_{s}) \\ \dot{G}_{0;1}(t,\boldsymbol{\lambda}_{s}) \\
\end{array}
\right) =
\hat{{\cal W}}(\boldsymbol{\lambda}_{s})
\left(
\begin{array}{c}
G_{1;1}(t,\boldsymbol{\lambda}_{s}) \\ G_{0;0}(t,\boldsymbol{\lambda}_{s}) \\
G_{1;0}(t,\boldsymbol{\lambda}_{s}) \\ G_{0;1}(t,\boldsymbol{\lambda}_{s}) \\
\end{array}
\right) \;,
\label{singleAaaa5}
\end{eqnarray}
where
\begin{eqnarray}
\hat{{\cal W}}(\boldsymbol{\lambda}_{s}) =
2 \left(
\begin{array}{cccc}
- v_{ss} & e^{\lambda_{(2,s)}} w_{ss}^{(L)}
+ e^{\lambda_{(4,s)}} w_{ss}^{(R)} & 0 & 0\\
e^{\lambda_{(1,s)}} v_{ss}^{(L)}
+ e^{\lambda_{(3,s)}} v_{ss}^{(R)} & -w_{ss} 
& 0 & 0\\
0 & 0 & -i \epsilon_s - v_{ss} - w_{ss} & 0 \\
0 & 0 & 0 & i \epsilon_s - v_{ss} - w_{ss}
\end{array}
\right) \;.
\label{singleAaaa5b}
\end{eqnarray}
Since we work in the reduced Liouville space (where FSC are neglected), 
only the coherences such that $\sum_{s=1}^{M} n_s = \sum_{s=1}^{M} n_s'$ 
are kept in (\ref{singleAaaa4}). \\

The two eigenvalues of the generator (\ref{singleAaaa5b}) 
corresponding to the population dynamics are given by
\begin{eqnarray}
g_{\pm}(\boldsymbol{\lambda}_{s}) &=&
-\left(\frac{v_{ss}+w_{ss}}{2}\right) \pm
\sqrt{\left(\frac{v_{ss}+w_{ss}}{2} \right)^2
+ f(\boldsymbol{\lambda}_{s})} \;, \label{singleAaaa6}
\end{eqnarray}
where
\begin{eqnarray}
f(\boldsymbol{\lambda}_{s}) &=& v_{ss}^{(L)}
\{ (e^{(\lambda_{(1,s)}+\lambda_{(2,s)})}-1) w_{ss}^{(L)}
+ (e^{(\lambda_{(1,s)}+\lambda_{(4,s)})}-1) w_{ss}^{(R)} \}
\nonumber\\
&&+ v_{ss}^{(R)} \{ (e^{(\lambda_{(2,s)}+\lambda_{(3,s)})}-1)
w_{ss}^{(L)}
+ (e^{(\lambda_{(3,s)}+\lambda_{(4,s)})}-1)
w_{ss}^{(R)} \} \;.\nonumber
\end{eqnarray}
The two eigenvalues corresponding to coherences are obviously 
given by the two lower diagonal elements of (\ref{singleAaaa5b})
since population are decoupled from coherences.
The right [left] eigenvectors of the generator 
can be easily evaluated. The two associated to the 
population read
$\hat{{\cal W}}_s (\boldsymbol{\lambda}_{s}) 
\sket{g_{\pm}(\boldsymbol{\lambda}_{s})}
= g_{\pm}(\boldsymbol{\lambda}_{s}) 
\sket{g_{\pm}(\boldsymbol{\lambda}_{s})}$
[$\sbra{\tilde{g}_{\pm}(\boldsymbol{\lambda}_{s})} 
\hat{{\cal W}}_s(\boldsymbol{\lambda}_{s}) 
=\sbra{\tilde{g}_{\pm}(\boldsymbol{\lambda}_{s})} 
g_{\pm}(\boldsymbol{\lambda}_{s})$].
By tracing the solution of (\ref{singleAaaa5}) 
we get the orbital GF 
\begin{eqnarray}
G_s(t,\boldsymbol{\lambda}_{s}) = 
c_+(0) e^{g_{+}(\boldsymbol{\lambda}_{s}) t}
\big( g^{+}_{1;1}(\boldsymbol{\lambda}_{s}) +
g^{+}_{0;0}(\boldsymbol{\lambda}_{s}) \big)
+ c_{-}(0) e^{g_{-}(\boldsymbol{\lambda}_{s})t}
\big( g^{-}_{1;1}(\boldsymbol{\lambda}_{s}) +
g^{-}_{0;0}(\boldsymbol{\lambda}_{s}) \big) \;,
\label{singleAaaa6a}
\end{eqnarray}
where $g^{\pm}_{n_s;n_s}(\boldsymbol{\lambda}_{s})=
\sbraket{n_s;n_s}{g_{\pm}(\boldsymbol{\lambda}_{s})}$ 
and 
$c_{\pm}(0)=\sbraket{\tilde{g}_{\pm}
(\boldsymbol{\lambda}_{s})}{G_s(0,\boldsymbol{\lambda}_{s})}$.
%The solution of (\ref{singleAaaa5}) therefore reads,
%\begin{eqnarray}
%\hspace{-1.5cm}
%\left(
%\begin{array}{c}
%G_{1}^{(s)}(t) \\ G_{0}^{(s)}(t)
%\end{array}
%\right) &=&
%c_+(0) e^{g_{+}(\boldsymbol{\lambda}_{s})t}
%\left(
%\begin{array}{c}
%g^{+}_{1}(\boldsymbol{\lambda}_{s}) \\ g^{+}_{0}(\boldsymbol{\lambda}_{s})
%\end{array}
%\right) +
%c_-(0) e^{g_{-}(\boldsymbol{\lambda}_{s})t}
%\left(
%\begin{array}{c}
%g^{-}_{1}(\boldsymbol{\lambda}_{s}) \\ g^{-}_{0}(\boldsymbol{\lambda}_{s})
%\end{array}
%\right) \;.
%\label{singleAaaa6a}
%\end{eqnarray}
%The
%generator $\hat{{\cal W}}(\boldsymbol{\lambda})$ does not couple the part of the
%reduced Liouville space corresponding to population with the part
%corresponding to coherences which is purely diagonal. Therefore, in
%taking the trace of the GO in order to get the GF [Eq.
%(\ref{ecountAaaarca})], out of $N_{\rm red}$ eigenvalues of the
%generator, only $2^M$ eigenvalues corresponding to the populations
%contribute. 
The many-body GF (\ref{ecountAaaatc}) is given by
\begin{eqnarray}
G(t,\boldsymbol{\lambda}) = 
\prod_{s=1}^{M} G_s(t,\boldsymbol{\lambda}_{s})\;.
\label{singleAaaa7}
\end{eqnarray}
This constitutes the solution of Eq. (\ref{ecountAaaatab}).
If one does the spectral decomposition directly on the many-body generator
one gets  
\begin{eqnarray}
G(t,\boldsymbol{\lambda}) = 
\sum_{m}^{2^M} e^{g_{m}(\boldsymbol{\lambda}) t}
\sbraket{I}{g_{m}(\boldsymbol{\lambda})} 
\sbraket{\tilde{g}_m(\boldsymbol{\lambda})}{\rho(0)}\;,
\label{singleAaaa7b}
\end{eqnarray}
where $g_{m}(\boldsymbol{\lambda})$, $\sket{g_{m}(\boldsymbol{\lambda})}$ and
$\sbra{\tilde{g}_{m}(\boldsymbol{\lambda})}$ are respectively the many-body
eigenvalues, right and left eigenvector of the generator. 
Each of these many-body eigenvalue is made from one of the $2^M$ 
possible ways of summing the orbital eigenvalues (\ref{singleAaaa6})
and the many-body left and right eigenvectors are tensor products of the 
single orbital eigenvectors.

%%%%%%%%%%%%%%%%%%%%%%%%%%%%%%%%%%%%%%%%%%%%%%%%%%%%%%%%%%%%%%%%%%%%%%%%%%%%%%%%%%%%%%%%%%%%%%%%%%
\section{Fluctuation theorem derived from the generating function symmetry} 
\label{largedev}

The following reasoning is based on the steady state FT for 
the entropy first obtained \cite{Lebowitz,Gaspard1} and 
later extended for currents \cite{GaspardAndrieux,GaspardAndrieux1}.
The GF is associated with the probability distribution by
\begin{eqnarray}
G(t,\lambda) = \sum_k P(t,k) e^{- \lambda k}
= \int d\xi \tilde{P}(t,\xi) e^{- \lambda \xi t} \;,
\label{FTaaaaaa}
\end{eqnarray}
where we have introduced $\tilde{P}(t,\xi)$, the 
probability that $\xi=k/t$ takes a value in the interval 
$[\xi,\xi+d\xi]$. 
The large deviation function (LDF) is defined as
\begin{eqnarray}
R(\xi) \equiv 
- \lim_{t \to \infty} \frac{1}{t} \ln \tilde{P}(t,\xi) \;.
\label{FTaaaaab}
\end{eqnarray}
This definition follows from the ansatz
\begin{eqnarray}
\tilde{P}(t,\xi) = C(\xi,t) e^{-R(\xi) t} \;,
\label{FTaaaaac}
\end{eqnarray}
where
\begin{eqnarray}
\lim_{t \to \infty} \frac{1}{t} \ln C(\xi,t) = 0 \;.
\label{FTaaaaaca}
\end{eqnarray}
We can then rewrite (\ref{FTaaaaaa}) as
\begin{eqnarray}
G(t,\lambda)= \int d\xi C(\xi,t) e^{- (R(\xi)+\lambda \xi) t}
\;. \label{FTaaaaad}
\end{eqnarray}
At long times, the main contribution to this integral comes 
from the value of $\xi$, $\xi^*$, that maximizes 
the argument of the exponential. 
$\xi^*$ is therefore the value of $\xi$ such that 
$\lambda=-\frac{dR}{d\xi} \vert_{\xi=\xi^*}$. 
At long times, using steepest descent integration, 
(\ref{FTaaaaad}) becomes
\begin{eqnarray}
G(t,\lambda) \approx e^{- (R(\xi^*)+\lambda \xi^*) t} \int d\xi 
C(\xi,t) e^{- \frac{1}{2} \frac{d^2 R(\xi)}{d \xi^2}\vert_{\xi=\xi^*} 
(\xi-\xi^*)^2 t} 
\approx e^{- (R(\xi^*)+\lambda \xi^*) t} C(\xi^*,t) 
\bigg( ( \frac{d^2 R(\xi)}{d \xi^2}\vert_{\xi=\xi^*} ) 
\frac{t}{2\pi} \bigg)^{-\frac{1}{2}} \;.
\label{FTaaaaadbis}
\end{eqnarray}
Substituting (\ref{FTaaaaadbis}) in (\ref{FTaaaaae}) gives
\begin{eqnarray}
S(\lambda) = R(\xi^*) + \lambda \xi^* \;. \label{FTaaaaaebis}
\end{eqnarray}
This shows that $S(\lambda)$ is the Legendre transform of the LDF.
The LDF is given by the inverse Legendre transform of $S(\lambda)$
\begin{eqnarray}
R(\xi) = S(\lambda^*) - \lambda^* \xi
\;, \label{FTaaaaaf}
\end{eqnarray}
where $\xi= \frac{d S}{d \lambda} \vert_{\lambda=\lambda^*}$.
Since $S(\lambda)$ is convex downward 
[i.e. $d^2 S / d \lambda^2=- \lim_{t \to \infty} 
(\langle k^2 \rangle_t - \langle k \rangle^2_t) \le 0$], 
its Legendre transform is convex upwards.
Using the symmetry (\ref{Caaaaib}), Eq. (\ref{FTaaaaaf}) implies that
$R(-\xi) = S(\beta e V - \lambda) + (\beta e V - \lambda) \xi$,
which together with $R(\xi) = S(\lambda) - \lambda \xi$ leads to
\begin{eqnarray}
R(\xi) - R(-\xi) = - \beta e V \xi \;.
\label{FTaaaaag}
\end{eqnarray}
Substituting this in Eq. (\ref{FTaaaaac}), we get
\begin{eqnarray}
\ln \frac{\tilde{P}(t,\xi)}{\tilde{P}(t,-\xi)}
=  \beta e V \xi t + \ln \frac{C(\xi,t)}{C(-\xi,t)} \;.
\label{FTaaaaaha}
\end{eqnarray}
Using (\ref{FTaaaaaca}), this gives the FT (\ref{FTaaaaah})
in the long time limit.

%%%%%%%%%%%%%%%%%%%%%%%%%%%%%%%%%%%%%%%%%%%%%%%%%%%%%%%%%%%%%%%%%%%%%%%%%
\section{Current, power spectrum and the generating function}
\label{momentsappendix}

We show how to calculate the average currents, their zero frequency 
power spectra and their Mandel parameter from the GF. 
These results are used in section \ref{AverageGF}. \\

To simplify the notation, we assume we have a probability 
distribution $P(t,\bold{k})$ and its associated generating function 
$G(t,\boldsymbol{\lambda})=\sum_{\bold{k}} P(t,\bold{k}) 
e^{\boldsymbol{\lambda} \cdot \bold{k}}$, where the component of
$\bold{k}$ and $\boldsymbol{\lambda}$ are given by $k_{\eta}$ 
and $\lambda_{\eta}$.
We also have $S(\boldsymbol{\lambda})=-\lim_{t \to \infty} 
\frac{1}{t} \ln G(t,\boldsymbol{\lambda})$.
The averaged number of charge $\eta$ is given by
\begin{eqnarray}
\mean{k_{\eta}}_t = - \frac{1}{e} \int_{0}^{t} d\tau \mean{I_{\eta}(\tau)}
= \left. \frac{\partial}{\partial \lambda_{\eta}}
G(t,\boldsymbol{\lambda}) \right|_{\boldsymbol{\lambda}=0} 
\;, \label{momentAaaaaa}
\end{eqnarray}
and the steady state current by 
\begin{eqnarray}
\mean{I_{\eta}}_{\rm st} &=& \lim_{t \to \infty} 
\frac{1}{t} \int_{0}^{t} d\tau \mean{I_{\eta}(\tau)} 
= \left. e \frac{\partial}{\partial \lambda_{\eta}}
S(\boldsymbol{\lambda}) \right|_{\boldsymbol{\lambda}=0}
\;. \label{momentAaaaaab}
\end{eqnarray}
We also find that
\begin{eqnarray}
\mean{k_{\eta} k_{\eta'}}_t =
\left. \frac{\partial}{\partial \lambda_{\eta}} \frac{\partial}{\partial \lambda_{\eta'}}
G(t,\boldsymbol{\lambda}) \right|_{\boldsymbol{\lambda}=0} = 
\frac{1}{e^2} \int_{0}^{t} d\tau_1 \int_{0}^{t} d\tau_2 \;
\mean{I_{\eta}(\tau_1) I_{\eta'}(\tau_{2})} \;.  \label{momentAaaaab}
\end{eqnarray}
Since at steady state
$\mean{I_{\eta}(\tau_1) I_{\eta'}(\tau_2)}_{st} =
\mean{I_{\eta}(\tau_1-\tau_2) I_{\eta'}(0)}_{st}$,
$\mean{I_{\eta}(\tau_1)}_{st}=\mean{I_{\eta}}_{st}$ and
$\mean{I_{\eta'}(\tau_2)}_{st}=\mean{I_{\eta'}}_{st}$,
using the fact that
\begin{eqnarray}
\left. \frac{\partial}{\partial \lambda_{\eta}} 
\frac{\partial}{\partial \lambda_{\eta'}}
\ln G(t,\boldsymbol{\lambda}) \right|_{\boldsymbol{\lambda}=0}
&=& \mean{k_{\eta} k_{\eta'}}_t - \mean{k_{\eta}}_t \mean{k_{\eta'}}_t \;,
\label{momentAaaaad}
\end{eqnarray}
we find
\begin{eqnarray}
A_{\eta \eta'} = e^2
\left. \frac{\partial}{\partial \lambda_{\eta}} \frac{\partial}{\partial \lambda_{\eta'}}
S(\boldsymbol{\lambda}) \right|_{\boldsymbol{\lambda}=0}
&=&\int_{-\infty}^{\infty} d\tau \bigg( \mean{I_{\eta}(\tau) I_{\eta'}(0)}_{st}
- \mean{I_{\eta}}_{st} \mean{I_{\eta'}}_{st} \bigg) \nonumber\\
&=&\int_{-\infty}^{\infty} d\tau \mean{[I_{\eta}(\tau) - \mean{I_{\eta}}_{st}]
[I_{\eta'}(0) - \mean{I_{\eta'}}_{st}]}_{st} \;.
\label{momentAaaaae}
\end{eqnarray}
Since the Fourier transform of the current correlation function is given by
\begin{eqnarray}
A_{\eta \eta'}(\omega) = \int_{-\infty}^{\infty} d\tau e^{- i \omega \tau}
\mean{[I_{\eta}(\tau) - \mean{I_{\eta}}_{st}] [I_{\eta'}(0) - \mean{I_{\eta'}}_{st}]}_{st} \;,
\label{momentAaaaaf}
\end{eqnarray}
we see that $A_{\eta \eta'}$ in (\ref{momentAaaaae}) is the zero frequency 
power spectrum of the current correlation function   
$A_{\eta \eta'} = A_{\eta \eta'}(\omega=0)$.
This quantity is used to study shot noise \cite{Buttiker}.\\
The analogue of the Mandel parameter in photon counting statistic
\cite{BarkaiRev,Mukamel06} for the process $y$ is given by
\begin{eqnarray}
M_{\eta}(t) \equiv 
\frac{\left( \mean{k_{\eta}^2}_t -\mean{k_{\eta}}_t^2 \right)
-\mean{k_{\eta}}_t}{\mean{k_{\eta}}_t}
= \frac{\left. \frac{\partial^2}{\partial \lambda_{\eta}^2} 
\ln G(t,\boldsymbol{\lambda}) \right|_{\boldsymbol{\lambda}=0}}
{\left. \frac{\partial}{\partial \lambda_{\eta}} 
G(t,\boldsymbol{\lambda}) \right|_{\boldsymbol{\lambda}=0}} - 1\;.
\label{spectdecAaaaal}
\end{eqnarray}
The asymptotic value is given by
\begin{eqnarray}
M_{\eta}(\infty) 
= \frac{\left. \frac{\partial^2}{\partial \lambda_{\eta}^2} 
S(\boldsymbol{\lambda}) \right|_{\boldsymbol{\lambda}=0}}
{\left. \frac{\partial}{\partial \lambda_{\eta}} 
S(\boldsymbol{\lambda}) \right|_{\boldsymbol{\lambda}=0}} - 1\;.
\label{spectdecAaaaalbis}
\end{eqnarray}
For a Poisson process $M_{\eta}=0$.
The zero frequency power spectrum is related to the long time limit of the 
Mandel parameter by
\begin{eqnarray}
A_{\eta \eta} &=& e \mean{I_{\eta}}_{\rm st} \big( 1 + M_{\eta}(\infty) \big)
\label{spectdecAaaaam} \;.
\end{eqnarray}

%%%%%%%%%%%%%%%%%%%%%%%%%%%%%%%%%%%%%%%%%%%%%%%%%%%%%%%%%%%%%%%%%%%%%%%%%%%%%%%%%%%%%%%%%%%%%%%%%%
\section{Generator for the two quantum dot model} \label{twolevel}

We present the basic quantities needed in the reduced Liouville space to 
study the model of two quantum dot model presented in section \ref{CurrentGF}.\\
The orbital eigenbasis is denoted by $\{\ket{n_1,n_2}\}$, where $n_1$ ($n_2$) 
is the occupation number of the orbital $s=1$ ($s=2$). 
Using the notation $\bra{n_1,n_2} \rho \ket{n_1',n_2'} = \rho_{n_1 n_2;n_1' n_2'}$,
the density matrix in the reduced Liouville space is given by the vector
$\rho = ( \rho_{00;00}\;,\; \rho_{01;01}\;,\; \rho_{10;10}\;,\;
\rho_{11;11}\;,\; \rho_{10;01}\;,\; \rho_{01;10} )^{T}$.
The generator of our QME (\ref{QMEaaaaaa}) for this model in this basis reads
\begin{eqnarray}
%\hspace{-1.6cm}
\hat{{\cal M}}=\left(
\begin{array}{cccccc}
-2 (w_{11}+w_{22}) & 2 v_{22} & 2 v_{11} & 0 &0 & 0 \\
2 w_{22} & -2 ( v_{22}+w_{11}) & 0 &
2 v_{11} & 0 & 0 \\
2 w_{11} & 0 & -2 (v_{11}+w_{22}) &
2 v_{22} & 0 & 0 \\
0 & 2 w_{11} & 2 w_{22}
& -2 (v_{11}+v_{22}) & 0& 0 \\
0 & 0 & 0 & 0 & -{\cal X}& 0 \\
0 & 0 & 0 & 0 & 0 & -{\cal X}^* \;,
\end{array}
\right)
\label{generatorQME2dots}
\end{eqnarray}
where ${\cal X}=v_{11}+w_{11}+v_{22}+w_{22}
+i(\epsilon_{1}-\epsilon_{2})$.
As expected, the populations are decoupled from the coherences.
The generator is diagonal for the coherences and obeys a birth and
death master equation in the population space.
The generator for the GO evolution equation (\ref{ecountAaaatab})
is given by $\hat{{\cal W}}(\boldsymbol{\lambda})=$
\begin{eqnarray}
\hspace{-0.2cm}
\left(
\begin{array}{cccc}
-2 (w_{11}+w_{22})
& 2 (e^{\lambda_{(1,2)}} v_{22}^{(L)}
+ e^{\lambda_{(3,2)}} v_{22}^{(R)})
& 2 (e^{\lambda_{(1,1)}} v_{11}^{(L)}
+ e^{\lambda_{(3,1)}} v_{11}^{(R)}) & 0 \\
2 (e^{\lambda_{(2,2)}} w_{22}^{(L)}
+ e^{\lambda_{(4,2)}} w_{22}^{(R)})
& -2 ( v_{22}+w_{11}) & 0 &
2 (e^{\lambda_{(1,1)}} v_{11}^{(L)}
+ e^{\lambda_{(3,1)}} v_{11}^{(R)}) \\
2 (e^{\lambda_{(2,1)}} w_{11}^{(L)}
+ e^{\lambda_{(4,1)}} w_{11}^{(R)})
& 0 & -2 (v_{11}+w_{22}) &
2 (e^{\lambda_{(1,2)}} v_{22}^{(L)}
+ e^{\lambda_{(3,2)}} v_{22}^{(L)}) \\
0 & 2 (e^{\lambda_{(2,1)}} w_{11}^{(L)}
+ e^{\lambda_{(4,1)}} w_{11}^{(R)})
& 2 (e^{\lambda_{(2,2)}} w_{22}^{(L)}
+ e^{\lambda_{(4,2)}} w_{22}^{(L)})
& -2 (v_{11}+v_{22})
\end{array}
\right)
\end{eqnarray}
where the coherence part has been discarded since it is the
same as for the generator of the QME.

%%%%%%%%%%%%%%%%%%%%%%%%%%%%%%%%%%%%%%%%%%%%%%%%%%%%%%%%%%%%%%%%%%%%%%%%%%%%%%%%%%%%%%%

%%%%%%%%%%%%%%%%%%%%%%%%%%%%%%%%%%%%%%%%%%%%%%%%%%%%%%%%%%%%%%%%%%%%%%%%%%%%%%%%%%%%%%
%%%%%%%%%%%%%%%%%%%%%%%%%
%%%%%%%%%%%%%%%%%%%%%%%%%
%%%%%%%%%%%%%%%%%%%%%%%%%
%%%%%%%%%%%%%%%%%%%%%%%%%
%%%%%%%%%%%%%%%%%%%%%%%%%
%%%%%%%%%%%%%%%%%%%%%%%%%
%%%%%%%%%%%%%%%%%%%%%%%%%
%%%%%%%%%%%%%%%%%%%%%%%%%
%%%%%%%%%%%%%%%%%%%%%%%%%
%%%%%%%%%%%%%%%%%%%%%%%%%
%%%%%%%%%%%%%%%%%%%%%%%%%%%%%%%%%%%%%%%%%%%%%%%%%%%%%%%%%%%%%%%%%%%%%%%%%%%%%%%%%%%%%%%%%%%%%%%%%%%%%%%%%%
\end{document}